\documentclass[11pt,preprintnumbers,aps,amsmath,onecolumn,nofootinbib]{revtex4}

\usepackage{epsfig}
\usepackage{graphicx}

\begin{document}

\title{\Huge  Introduction –- Strong interaction 
in the nuclear medium:  new trends 
\vspace{.5 cm}}

\author{\Large Denis Lacroix}

\affiliation{GANIL, B.P. 55027,
F-14076 CAEN Cedex 5, France}


\maketitle

\vspace{.cm}

\centerline{\large ABSTRACT}

\vspace{.5 cm}

Recent achievements in nuclear forces theory open new perspectives for the next decade 
of low energy nuclear physics, bringing together people from very different communities. Although many developments 
remain to be done, the possibility to directly use QCD to describe nuclear system is a major challenge that is 
within reach. 
In this introduction to the 2009 International Joliot-Curie School (EJC2009), new trends in the
strong nuclear interaction are summarized starting from quarks and ending with finite or infinite nuclear 
systems. At different energy scales, selected new concepts and ideas have been discussed in a rather
simple way. Recent advances in the theory of nuclear forces, thanks to chiral perturbation  
and effective field theories, have led to a new generation of strong nuclear interaction particularly 
suited to low energy nuclear physics. The interesting aspects of new interactions compared to conventional forces 
are underlined. Recent achievements 
in ab initio theories that directly start from the bare nucleon-nucleon interaction and their key role to 
understand the three-body force are illustrated. Finally, future perspectives for standard nuclear physics 
theories, namely Shell Model and Energy Density Functional, are discussed.


\tableofcontents

\section{Introduction}

\begin{figure}[htb]
\begin{center}
\includegraphics[width = 16.cm]{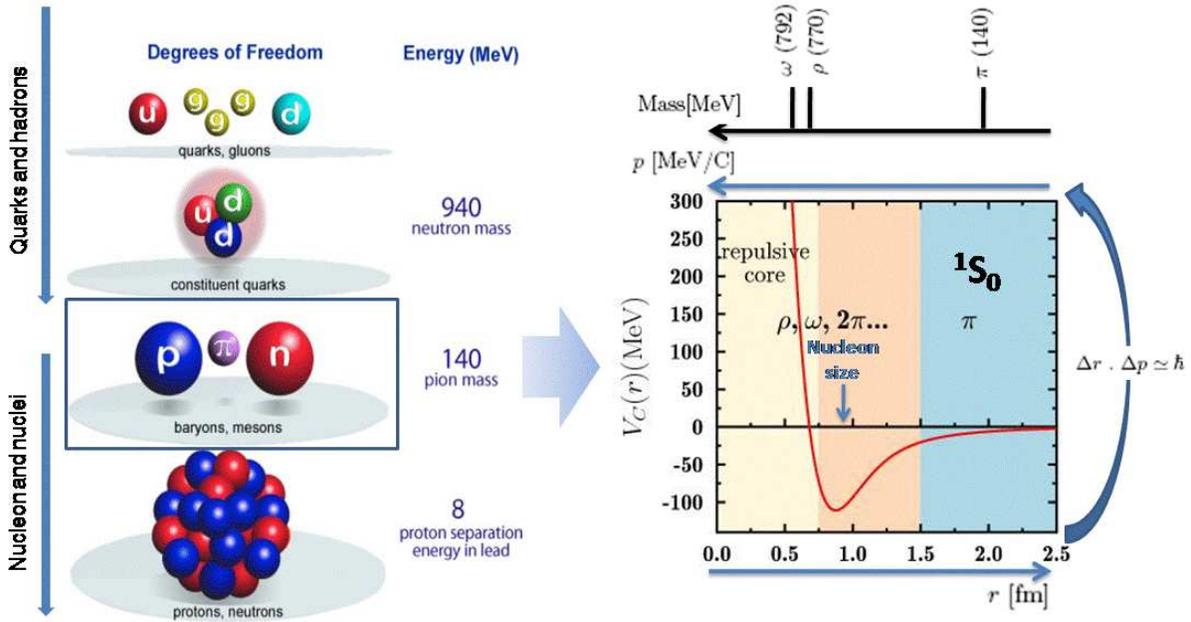}
\end{center}
\caption{(Color online) Left: The strong nuclear interaction is at the crossing point between  "High energy" nuclear physics 
dedicated to quarks and how they organize into nucleons and "Low energy" nuclear 
physics dedicated to nucleons and how they organize into nuclei. Right: Typical 
example of central nuclear interaction as a function of relative distance between the two nucleons. The relative distance 
can directly be related to the energy range involved in the interaction. Using the Heisenberg uncertainty, the typical momentum 
exchange at a given $r$ is given by $p \propto \hbar/(2 r)$ while the scale of energy involved is proportional to the square 
of $p$. A meson is involved in the interaction if its mass is in the energy scale associated to a given $r$. For instance, the
$\pi$ is the only meson that contributes to the  
long-range part of the interaction. At shorter relative distance, 
more and more mesons contribute to the interaction.}
\label{fig:intro1}
\end{figure}

Two fundamental questions of present days nuclear physics are (i) How to understand the very rich 
structure of atomic nuclei in terms of interaction between 
nucleons? (ii) How to relate the strong nuclear interaction to the underlying fundamental Chromodynamics (QCD)
that governs the physics of quarks and gluons. These two questions illustrate the many facets of nucleon-nucleon 
interactions (see Fig. \ref{fig:intro1}). "Low energy" nuclear scientists mainly address (i) and often 
consider the strong interaction as a "fundamental" interaction and nucleons as elementary (often point-like) particles.
From the "High energy" nuclear physics point of view, nucleons, being formed by quarks and gluons, 
can obviously  not be considered as elementary particles and the strong interaction itself should be understood 
from the more fundamental Standard Model. The nuclear force is at the crossroad of these two visions. 
Recently, important progresses have been made in the understanding of the nuclear strong interaction directly 
from QCD. Conjointly, new experimental results have pointed out our lack of knowledge of the interaction 
in the nuclear medium.  Selected issues directly related to the strong interaction and its effects 
are listed below: 
\begin{itemize}
\item Lattice QCD is a promising tool to calculate ab initio hadrons and mesons properties on a mesh. 
Recently, first applications of Lattice QCD have been made possible \cite{Ish07}. 
\item The use of Effective Field Theory in combination with Chiral perturbation theory 
provides a systematic and elegant 
framework to construct the nuclear forces for QCD Lagrangian \cite{Epe08}. 
\item New "soft" interactions, like the so-called $V_{\rm lowk}$, 
deduce from conventional or more recently similarity renormalization group technique
have been proposed \cite{Bog05}. These interactions open new perspectives 
by getting rid of the hard core while keeping the low energy 
nuclear physics unchanged. 
\item With recent soft- and hardware progresses, systematic ab initio calculations for 
light nuclei starting directly from the bare nucleon-nucleon interaction 
are now possible.  
\item Comparison of these ab initio theory with high precision measurements have confirmed and precised the 
importance of three-body forces. 
\item Specific components of the force (spin-orbit, tensor interaction, 3-body) have been largely debated
in the low energy nuclear physics community \cite{Sor08} underlying the necessity to use less phenomenological approaches.  
\item Standard theories of low energy nuclear physics, like the nuclear Shell Model \cite{Cau05} or the Energy Density 
Functional \cite{Ben03}, that generally start from interactions directly adjusted to reproduce data,  
are now being revisited using a "bottom-up" approach, i.e. starting from the QCD ingredients and ending with the nuclei
properties. 
\end{itemize} 
The aim of the 2009 Joliot-Curie international School is to present new advances in the comprehension and use of the 
nuclear interaction. It covers not only the physics of quarks, nucleons, nuclei and hypernuclei as well as stars. 
These subjects are certainly far too broad to expect that a single physicist that is often specialized in much 
narrower fields. However, it is essential that we learn enough from each others to be able to communicate 
and cross fertilize.
  
The list of new aspects quoted above will be extensively discussed in the different lectures of this 
school. The objectives of the present introduction are (i) To provide a roadmap for the school, that, I hope, will help the 
reader to make connection between different lectures (ii) To provide simple summary on the ongoing research. 
The price to pay is to use shortcuts and simplified pictures that  might appear too schematic for specialists. People 
interested in detailed discussions can directly read dedicated lectures of this school. 
(iii) to emphasize the new aspects that will impact the future of nuclear
physics. Being from the Low energy branch of nuclear physics, the overview given below is highly personal 
and obviously subjective.  
The present lecture is organized as follows: First, basic ingredients of conventional nuclear 
forces are recalled. Then, new perspectives offered by the Effective Field Theory and Chiral Perturbation 
Theory are discussed. Last, implications for the description and understanding of nuclei are underlined. 

\section{Conventional nuclear Forces}
\label{sec:conventional}

The aim of the present section is not to give a broad review on nuclear interaction that have been 
proposed in the past. An excellent review can be found for instance in \cite{Mac94,Mac01} as well as a complete 
description of conventional nuclear forces. Here, I recall some basic notions that will be 
helpful for the present discussion. 

Conventional forces generally start from semi-phenomenological parametrisations that could be written 
schematically as:
\begin{eqnarray}
V_{NN} &=& v^{\rm EM}_{NN} + v^{\rm \pi}_{NN} + v^{\rm Rep}_{NN} . 
\end{eqnarray}  
The first part denotes the Electromagnetic (EM) contribution. $v^{\rm \pi}_{NN}$ stands for the One-Pion Exchange 
potential responsible for the long-range attraction. $v^{\rm Rep}_{NN}$ corresponds to the strong repulsive short 
range interaction. It is generally taken as a phenomenological parametrization. Altogether, most recent 
nuclear interactions have around 30 parameters that are adjusted on experimental nucleon-nucleon 
scattering cross sections.

\subsection{Classification of nuclear forces}

\begin{figure}[htb]
\begin{center}
\includegraphics[width = 12.cm]{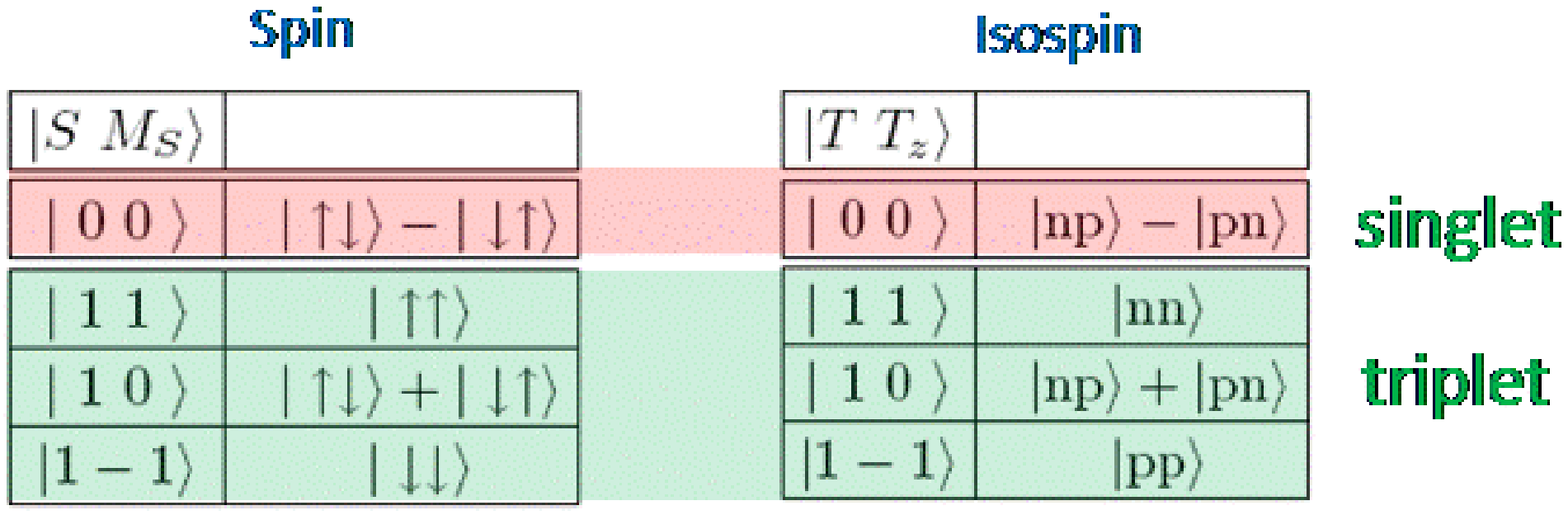}
\end{center}
\caption{(Color online) Illustration of the different projection of total spin (left) and total isospin (right)}
\label{fig:spinisospin}
\end{figure}

\begin{figure}[htb]
\begin{center}
\includegraphics[width = 11.cm]{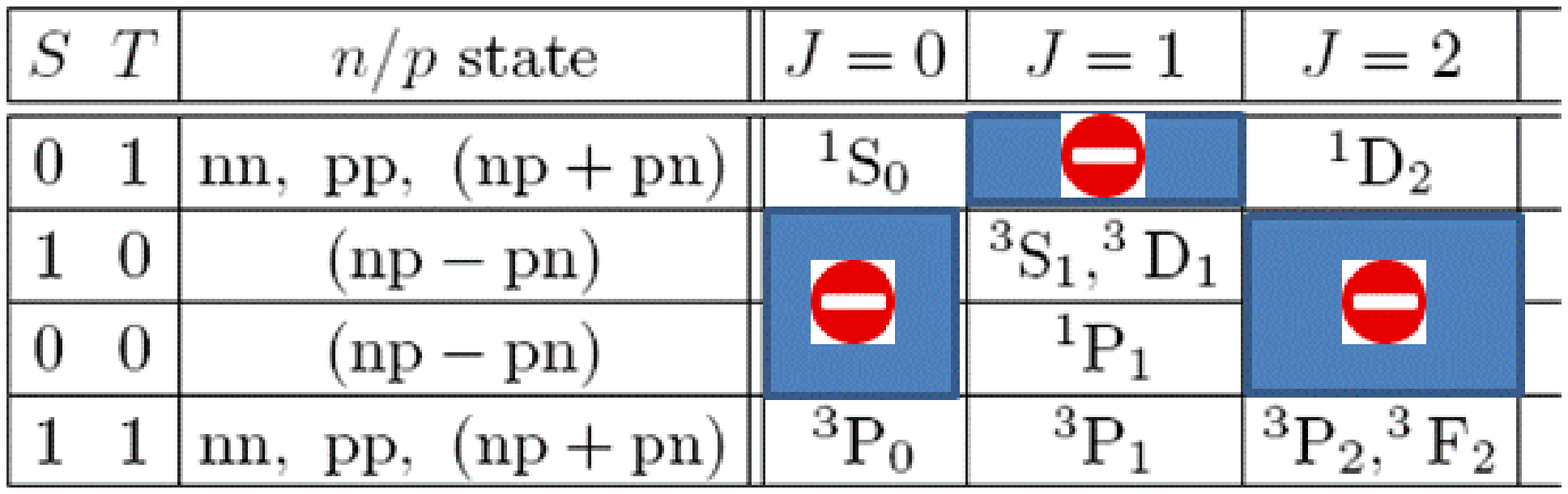}
\end{center}
\caption{(Color online) Illustration of the most common channels generally used for the nuclear interaction. The forbidden 
channels due to quantum selection rules are also indicated.}
\label{fig:channel}
\end{figure}

In figure \ref{fig:intro1}, the central part of the interaction in the $^1$S$_0$ channel is displayed. For non experts, the 
meaning of this denomination is recalled here. The properties of the nuclear interaction can be studied by introducing 
the eigenstates of the associated two-body problem, denoted here by 
$\Psi_{\rm NN}  = \Psi ( \mathbf{r}_1 \sigma_1 \tau_1 ; \mathbf{r}_2 \sigma_2 \tau_2 )$ where 
$\sigma$ ($\uparrow $ or $\downarrow$) and $\tau$ (n or p) 
denote respectively the spin and isospin quantum numbers. By coupling the two spins on one side and the two 
isospin on the other side, the total spin $S$ and isospin $T$ and associated projections $M_S$ and $T_z$ are generally 
introduced. As an illustration, the different components of the projections are given in figure \ref{fig:spinisospin}.
Since the nuclear interaction is spherical symmetric, the orbital angular momentum $L$ that couples to the 
spin $S$ as $J = L + S$
can be introduced, where $J$ is the total angular momentum. Then, a given channel of the interaction is denoted by:
\begin{eqnarray}
^{2S+1}\left[L \right]_{J} \hspace*{0.5cm} {\rm with }  \hspace*{0.5cm} 
\left[L=0,1,2, \cdots \right] =~{\rm S},~{\rm P},~{\rm D},\cdots  \nonumber
\end{eqnarray}
Some of the most common channels and their content in terms of isospin and total angular momentum 
are given in figure \ref{fig:channel}.
 
\subsection{Scattering theory and Phase-shift analysis}

\begin{figure}[htb]
\begin{center}
\includegraphics[width = 14.cm]{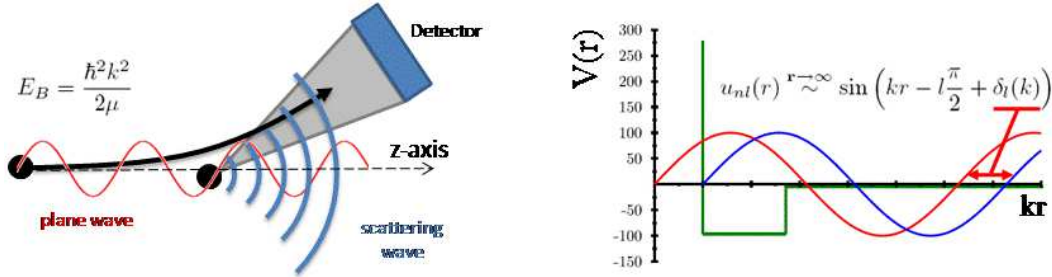}
\end{center}
\caption{(Color online) Left: Schematic illustration of the diffusion of one nucleon onto another nucleon 
with incident energy $E = \frac{\hbar k^2}{2\mu}$. 
Here, $\mu$ is the reduced mass of the collision. Right: Illustration of the effect of the interacting 
potential. Here a schematic potential is used, the red curve is the free nucleon case while the blue curve corresponds to 
the eigenstate with two-body interaction. The phase-shift $\delta_l (k)$ is directly related to the difference between 
the asymptotic behavior of the two curves. }
\label{fig:scatter}
\end{figure}

The best way to characterize the interaction between two particles is to make them collide and detect 
the product of the reaction (see illustration in Fig \ref{fig:scatter} (left)). 
In scattering theory, the cross section is best studied by introducing the eigenstates of the two-body problem, denoted 
by $\Psi_{\rm scat}$. 
Since detectors are positioned at almost infinite distance, the probability to detect a particle at a given 
position is only sensitive to the asymptotic behavior of the wave-function, i.e. (see for instance \cite{Coh77})   
\begin{eqnarray}
\Psi_{\rm scat}({\mathbf r}) \stackrel{{\mathbf r} \rightarrow \infty } {\longrightarrow}  e^{ikz} 
+ f(\theta, \varphi) \frac{e^{ikr}}{r} \nonumber 
\end{eqnarray} 
where we recognize the superposition of the incident wave-function of a particle with an energy $E =\frac{\hbar k^2}{2\mu} \displaystyle$ 
and a scattering wave. Then, the cross section detected at given angles $\theta$ and $\varphi$ (where these angles 
correspond to spherical coordinates, $z$ being the beam axis), is related to the scattering amplitude through the simple relation 
$\sigma_{k}(\theta, \varphi) = |f(\theta, \varphi)|^2$.   
Taking advantage of the nuclear interaction spherical symmetry, the scattering wave-packet can be further decomposed 
on eigenstates with good angular momentum as   
\begin{eqnarray}
\Psi_{\rm scat}({\mathbf r}) &=& \sum_{nlm} c_{nlm} \frac{u_{nl}(r)}{r} Y_{lm}(\theta, \varphi). \nonumber 
\end{eqnarray}
We are then left with the study of the asymptotic behavior of the $u_{nl}(r)$ that is usually 
written  as
\begin{eqnarray}
u_{nl}(r)  \stackrel{{\mathbf r} \rightarrow \infty } {\sim} \sin 
\left( kr - l \frac{\pi}{2} +\delta_l(k) \right) \nonumber
\end{eqnarray}
where $\delta_l(k)$  denotes the phase-shift at a given $l$ and energy (note that here for the sake of simplicity, spinless particles have been 
considered). The physical meaning of the phase-shift is illustrated in figure \ref{fig:scatter} (Right). As a final result, 
the cross-section integrated over the angles can be simply written as
\begin{eqnarray}
\sigma (k) &=& \frac{4\pi}{k^2} \sum_l (2l+1) \sin^2 [\delta_l(k)]. \nonumber
\end{eqnarray}

\begin{figure}[htb]
\begin{center}
\includegraphics[width = 14.cm]{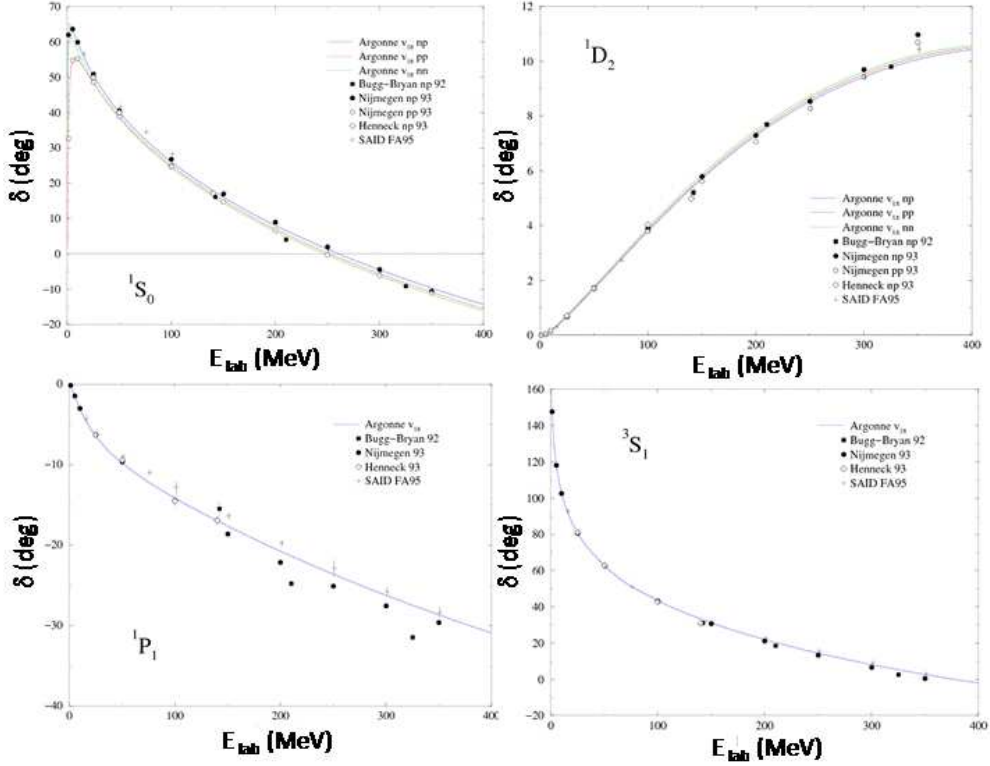}
\end{center}
\caption{(Color online) Illustration of a comparison between experimental and theoretical phase-shift as a function 
of laboratory colliding energy. The theoretical curves have been obtained with the AV18 interaction (Adapted from \cite{Wir95}).}
\label{fig:phase2}
\end{figure}

In practice, phase-shifts are extracted from experimental analysis of nucleon-nucleon cross sections at different energies. 
Then parameters of the nucleon-nucleon interaction are adjusted to best reproduce their behavior. 
An example of result obtained with a
conventional nuclear force, namely AV18 \cite{Wir95}, is shown in figure \ref{fig:phase2} for various channels. 
A detailed discussion on different refinement to adjust forces can be found in \cite{Mac01}.  
          
\subsection{Critical discussion}

Conventional nuclear forces have been adjusted for many years to reproduce scattering experiments and, most often, also 
deuteron properties. 
\begin{figure}[htb]
\begin{center}
\includegraphics[width = 7.cm]{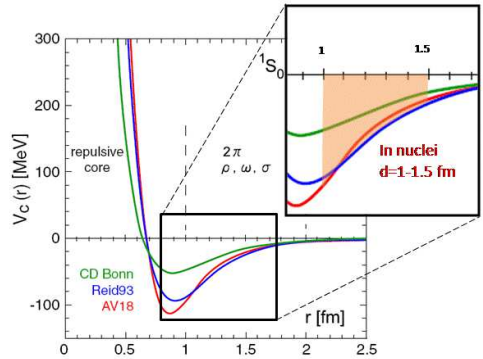}
\end{center}
\caption{(Color online) Comparison between three different conventional forces 
in the $^1$S$_0$ (spin singlet and s-wave) channel: CDBonn \cite{Mac01b}, Reid93 \cite{Sto93}, and AV18 \cite{Wir95}.
The insert shows the difference for scales comparable to the relative distance between nucleons inside nuclei.}
\label{fig:pot2}
\end{figure}
Several forces have been proposed that are dedicated to the same purpose. However, a direct 
comparison of these forces between each others shows significant differences. This is illustrated in figure \ref{fig:pot2}
where three different parameterizations are compared. These differences stem from different strategies used to design 
the forces themselves. One of the ambiguity of conventional forces is the absence of a constructive framework 
that would lead to a systematic improvement of the nuclear interaction and remove some of
the ambiguities. Such a constructive framework is provided by the Effective Field Theory combined with 
Chiral perturbation theory (see discussion below).         

\begin{figure}[htb]
\begin{center}
\includegraphics[width = 12.cm]{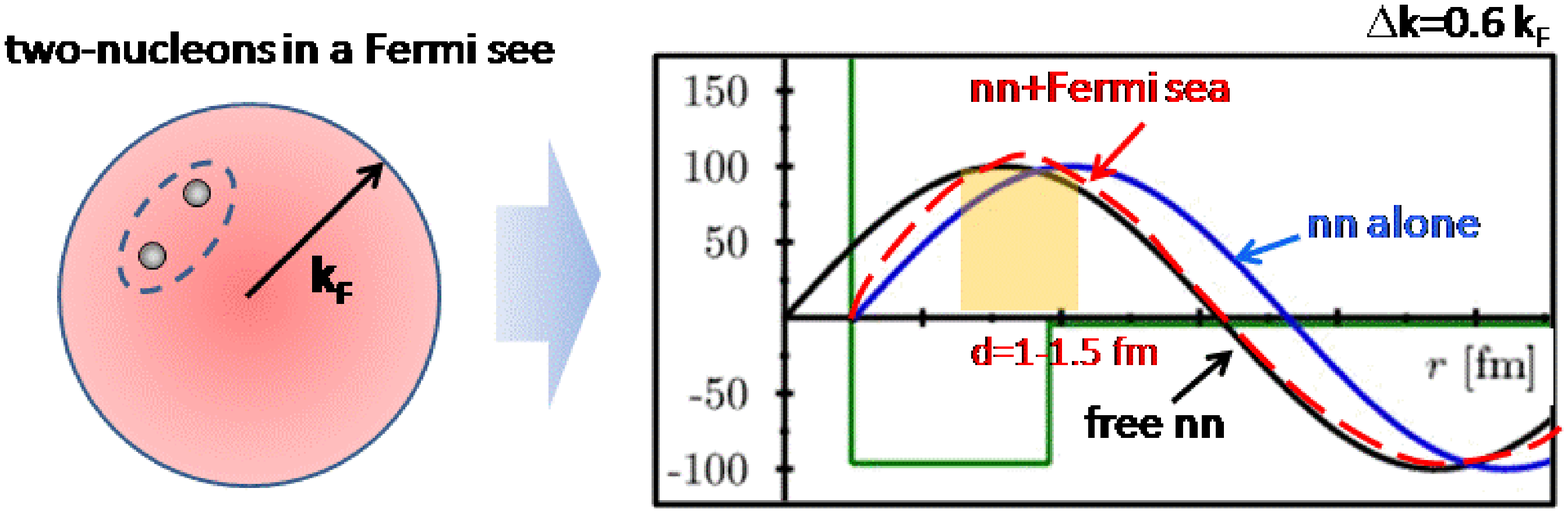}
\end{center}
\caption{(Color online) Left: Two nucleons plunged in a Fermi sea are considered. Only the interaction between the two 
nucleons is assumed to be non-zero, the surrounding nucleons affect the two nucleon sub-system through the Pauli principle 
by blocking accessible configurations. Left: a schematic nucleon-nucleon interaction is considered. The different 
curves correspond respectively to the two nucleon wave-function obtained for the 
free nucleon case (black solid line), i.e. without interaction and without the Fermi sea, 
the isolated two nucleon case (blue solid line), i.e. the two nucleon interacting through the bare interaction without 
the Fermi sea (this case is called "nn alone" in the figure), 
the two interacting nucleons plunged into the Fermi sea (green dashed line). As can be seen from 
the figure, contrary to the "nn alone" case, when the nucleons are plunged in a nuclear medium, the corresponding 
wave-function tends to the free nucleon case for $r>d$, where $d$ is known as the healing distance. }
\label{fig:medium} 
\end{figure}

A second difficult aspect, that is not specific to conventional nuclear forces, is related to in-medium effects. When 
two nucleons are in interaction inside a nucleus, the bare interaction is modified due to the presence of 
other surrounding nucleons. 
In figure \ref{fig:medium}, a schematic illustration of what are in-medium effects is given. 
When two nucleons are surrounded by other nucleons, the associated wave function looks very 
much like an independent particle case. This stems from a combined effects of the Pauli principle that
blocks accessible configurations for the two nucleons and the properties of 
the force itself. This simple analysis has been used for instance 
to reconcile on one hand the strength of the interaction and the independent particle picture validity in nuclei
(for further discussion see \cite{Gom58,Boh69}).
It can be conclude from figure \ref{fig:medium} 
that in-medium effects largely modify the interaction that is felt by nucleons 
compared to the original bare interaction and a proper dressing of the interaction is mandatory. Due to the 
repulsive core, specific methods based on the Bethe-Goldstone approximation have been developed \cite{Fet03}.   
These methods are rather involved and render the use of bare interactions in low energy nuclear model difficult.
As we will see below, the recent introduction of soft interactions provides a suitable solution to these 
difficulties.
 
\section{From quarks to nucleons: new aspects}
\label{sec:modernforce}

Modern theories of nuclear forces aim at developing nucleon-nucleon interaction directly from the underlying 
QCD Lagrangian, denoted hereafter by ${\cal L}_{\rm QCD}$. Here, basic but important notions on quarks and their 
organizations into mesons and hadrons are first recalled. Then, highlights on the different strategies used 
to derive NN interaction from QCD are given.

\subsection{Reminder on basic aspects on quarks, mesons and baryons and Lattice QCD}

\begin{figure}[htb]
\begin{center}
\includegraphics[width = 14.cm]{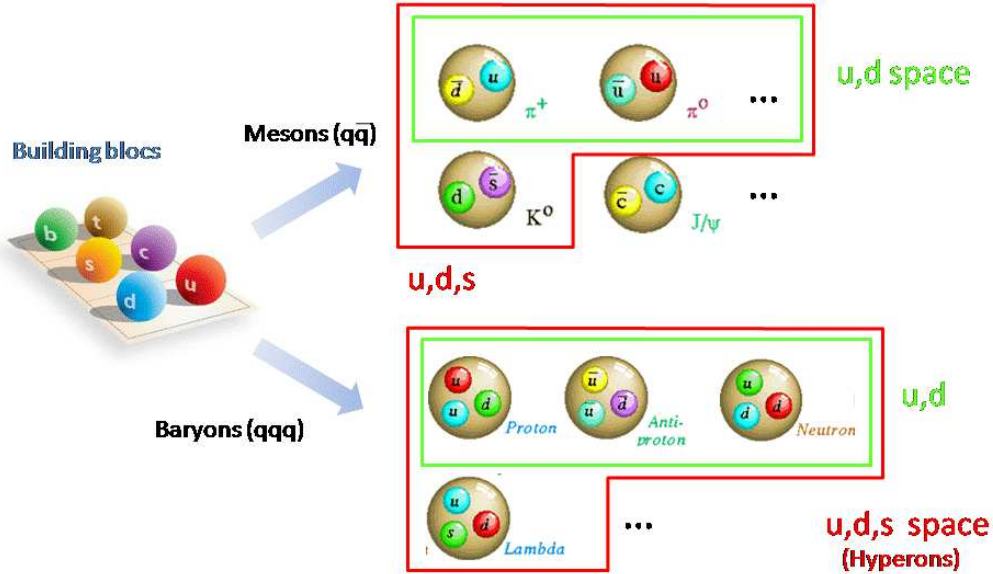}
\end{center}
\caption{(Color online) Illustration of the different quarks forming mesons and baryons. }
\label{fig:quark}
\end{figure}

Mesons and baryons are made from the aggregation of two or three quarks (or anti-quarks) respectively interacting through 
gluons. In figure \ref{fig:quark}, the constituents of different particles under interest are recalled. Low energy 
nuclear physics mainly focus on particles formed from quarks $u$ ($\bar u$) and $d$ ($\bar d$) forming for instance $\pi$, 
neutrons and protons. In Hyperons, that will be discussed below, one of the up or down quarks is replaced by a $s$ quark.     

\begin{figure}[htb]
\begin{center}
\includegraphics[width = 7.cm]{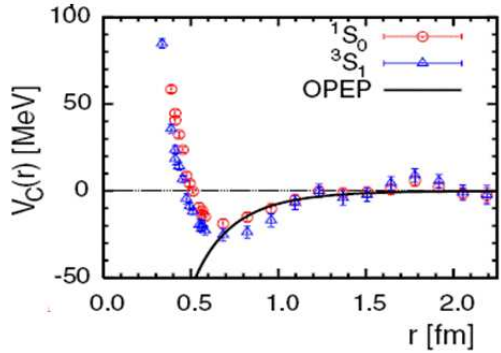}
\end{center}
\caption{(Color online) The lattice QCD result of the central
(effective central) part of the NN potential in the 
$^{1}$S$_0$ and $^{3}$S$_1$ channel. The solid lines correspond to the one-pion exchange
potential (OPEP) \cite{Ish07}.}
\label{fig:potquark}
\end{figure}

The QCD Lagrangian writes as:
\begin{eqnarray}
{\cal L}_{\rm QCD} &=& {\cal L}^0_{\rm QCD} - \bar q M q , \label{eq:lagqcd}
\end{eqnarray}    
where the first part corresponds to the Lagrangian with massless quarks 
while the second part stands for the mass contribution. 
As far as the strong interaction is concerned, the different quarks $(u, d, s)$ have
identical properties, except for their masses. The quark
masses are free parameters in QCD, i.e. the theory can be
formulated for any value of the quark masses. One of the difficult aspect 
of QCD physics is that it is highly non-perturbative in the low energy regime. 
For many years, Lattice QCD approach, where quarks and gluons fields are 
considered on a discretized mesh \cite{Dav02,Deg06}, has been though as the only framework 
suitable for this problem. Lattice QCD calculations require large 
computational power. As a consequence, first lattice QCD calculation 
of the NN interaction have only recently been performed \cite{Ish07} (see figure \ref{fig:potquark}). 
Lattice QCD application are numerical rather involved and approximations are still necessary 
to be able to obtain NN interaction like the one displayed in figure \ref{fig:potquark}. 

\subsection{Quark masses, Chiral symmetry and Effective Field Theory}

The introduction of Effective Field Theory has open an alternative to Lattice QCD 
to derive nucleon-nucleon interaction from the QCD Lagrangian.  
To understand recent developments on 
nuclear forces it is crucial to keep in mind the order of magnitude of quark masses:
\begin{eqnarray}
\left.
\begin{array} {l}
m_u  \sim  2.4^{\pm 0.9}{\rm MeV} \\
m_d  \sim  4.75^{\pm 1.25} {\rm MeV} \\ 
m_s  \sim  100^{\pm 30} {\rm MeV}
\end{array}
\right|  
\begin{array} {c}
{\rm Energy} \\
{\rm GAP}
\sim 1~{\rm GeV}
\end{array}
\left|
\begin{array} {l}
m_c \sim  1270^{+70}_{-110}{\rm MeV}  \\
m_b \sim  4200^{+170}_{-70} {\rm MeV} \\
m_t \sim  171200^{\pm 2100} {\rm MeV}
\end{array} \right.
\label{eq:massquarks}
.
\end{eqnarray}
Several important comments can be made from these values: (i) Masses of  
$(u, d, s)$ (called light quarks) are rather small even compared to the nucleon mass (ii) There exist 
a gap in energy around $1$ GeV between these quarks and the $(c,b,t)$ quarks (called Heavy quarks).

\subsubsection{Chiral symmetry breaking and its consequences}

\begin{figure}[htb]
\begin{center}
\includegraphics[width = 16.cm]{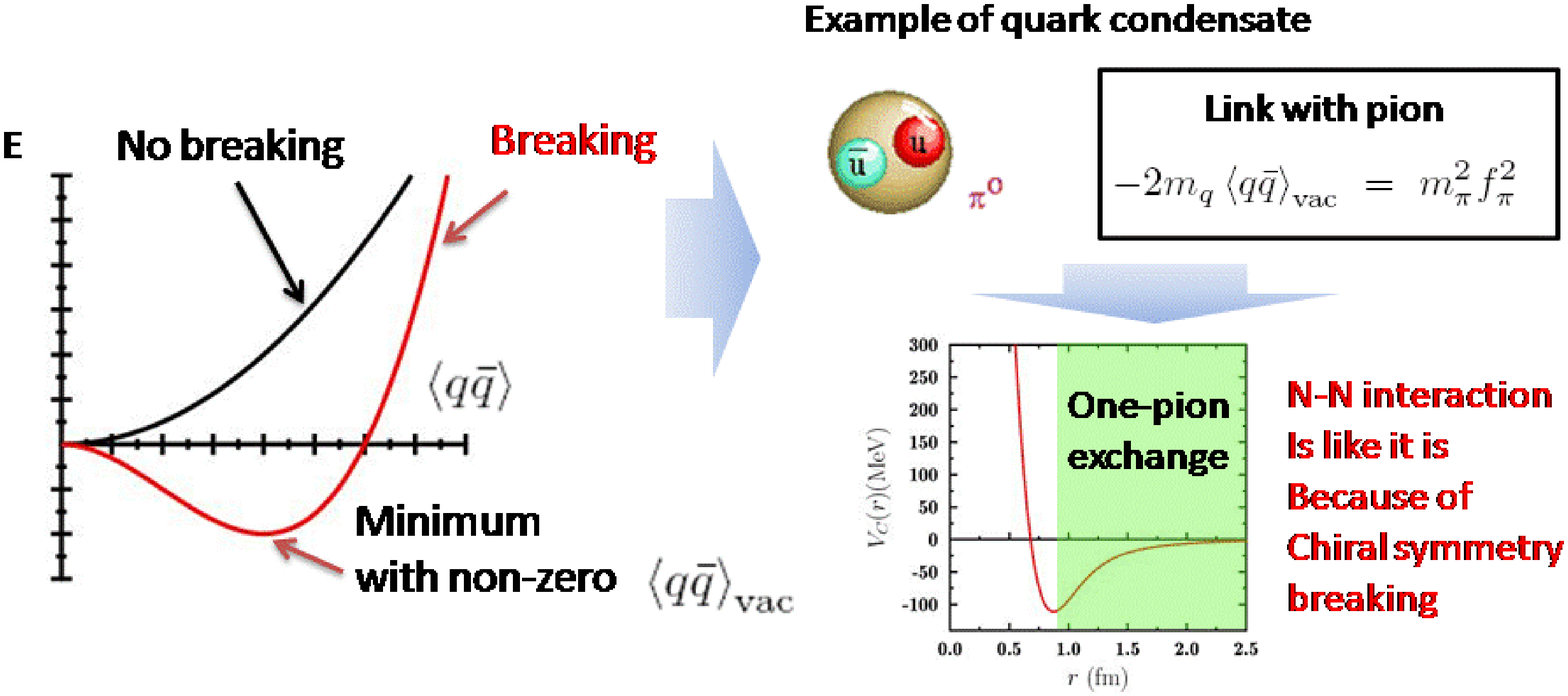}
\end{center}
\caption{(Color online) Left: Schematic illustration of the chiral symmetry breaking effect. As always, when a symmetry is 
broken one could defined an order parameter, here the quark-antiquark condensate $\langle q \bar q \rangle$. 
Without symmetry breaking, the energy landscape (here shown in one dimension for simplicity) 
has a minimum for zero value of the order parameter. When the symmetry is broken 
a minimum appear for non-zero value of $\langle q \bar q \rangle$. Right: the value of $\langle q \bar q \rangle$
at the minimum is directly related to the pion mass which is itself the main actor contributing to the 
long-range part of the strong interaction.}
\label{fig:chiral}
\end{figure}

Let us first, for simplicity, neglect the effect of $(c,b,t)$ quarks and focus on 
the subspace $(u,d,s)$ which is a priori relevant for low energy nuclear physics. Assuming 
that quarks are massless, the second part in Lagrangian (\ref{eq:lagqcd}) cancels out. 
One important properties of ${\cal L}^0_{\rm QCD}$ is that it is invariant under Chiral symmetry.
Because quarks have masses, the Chiral symmetry is explicitly broken, i.e. chiral   
transformation induces a mixing of Left-Handed (LH) and Right-Handed (RH) quarks. Chiral symmetry 
breaking plays a special role in the properties of nuclear forces, see illustration in figure (\ref{fig:chiral}).      
Chiral Perturbation Theory takes advantage of the smallness of $(u,d,s)$ masses by treating the second term 
in equation (\ref{eq:lagqcd}) directly as a perturbation.

\subsubsection{Effective Field Theory and Soft interactions} 

In the discussion above, $(c,b,t)$ quarks have been completely disregarded. The main argument to do so 
is that the energy scale under interest is much smaller than the typical energy of these quarks. Effective 
Field Theory provides a framework to properly disregard some degrees of freedom when two separated 
scales (here in energy) coexist in a physical problem. 

\begin{figure}[htb]
\begin{center}
\includegraphics[width = 16.cm]{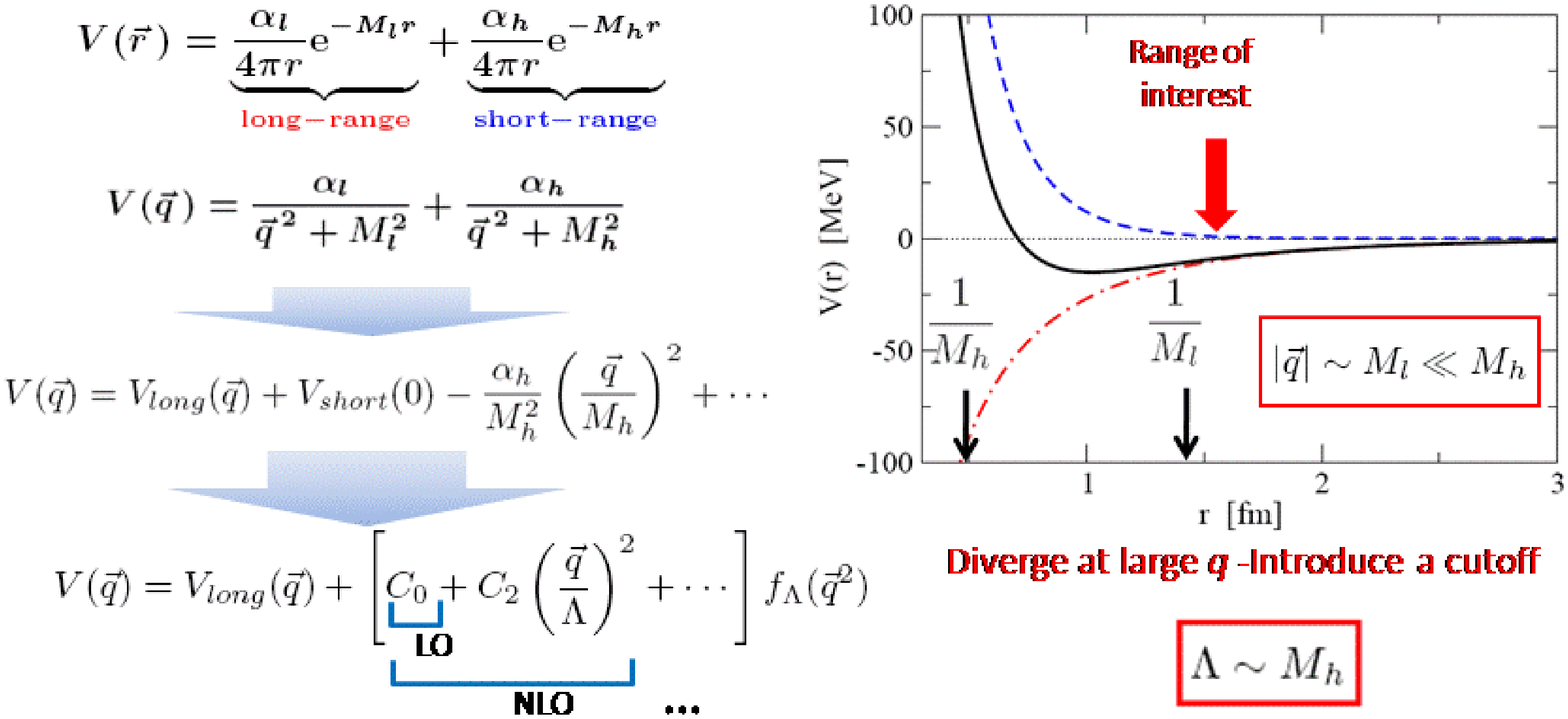}
\end{center}
\caption{(Color online) Illustration of the different steps used in Effective Field Theory on a schematic interaction (taken 
from ref. \cite{Mei08}). Here, the interaction (shown in the right side with solid line) 
writes as a sum of a long-range (dot-dashed line) and a short-range (dashed line) interaction respectively 
associated with the energy scale $M_l$ and $M_h$. Within EFT framework, the short range contribution is expanded 
in powers of $({\vec q} / M_h)$, where ${\vec q}$ denotes the momentum transfer during the nucleon-nucleon 
interaction. Such an expansion is expected to be valid at small ${\vec q}$ compared to the typical energy scale 
associated to the short-range interaction. Direct Taylor expansion of $V_{\rm short}$ presents a ultra-violet 
divergence as ${\vec q}$ increases. To avoid this divergence, a smooth function $f_\Lambda$, associated to a cut-off 
parameter $\Lambda$ is introduced, where $\Lambda \sim M_h$ . Often, the expansion is directly made as a function 
of  $({\vec q} / \Lambda)$. The accuracy of the expansion depends of the expansion order, namely Leading Order (LO), 
Next to Leading Order (NLO) ...}
\label{fig:eft1}
\end{figure}

The idea behind EFT is to keep unchanged the physic at the energy scale under interest while providing more and more 
accurate approximations for the high energy scale \cite{Wei91,Van99}. The strategy used is depicted in figure \ref{fig:eft1} for a toy model 
introduced by Meissner in \cite{Mei08}. 
As a result of the EFT, the interaction with a repulsive hard-core is replaced by a smooth interaction 
while the physics within a given range of energy, typically ${\vec q} << \Lambda$, is left 
unchanged.     

\begin{figure}[htb]
\begin{center}
\includegraphics[width = 14.cm]{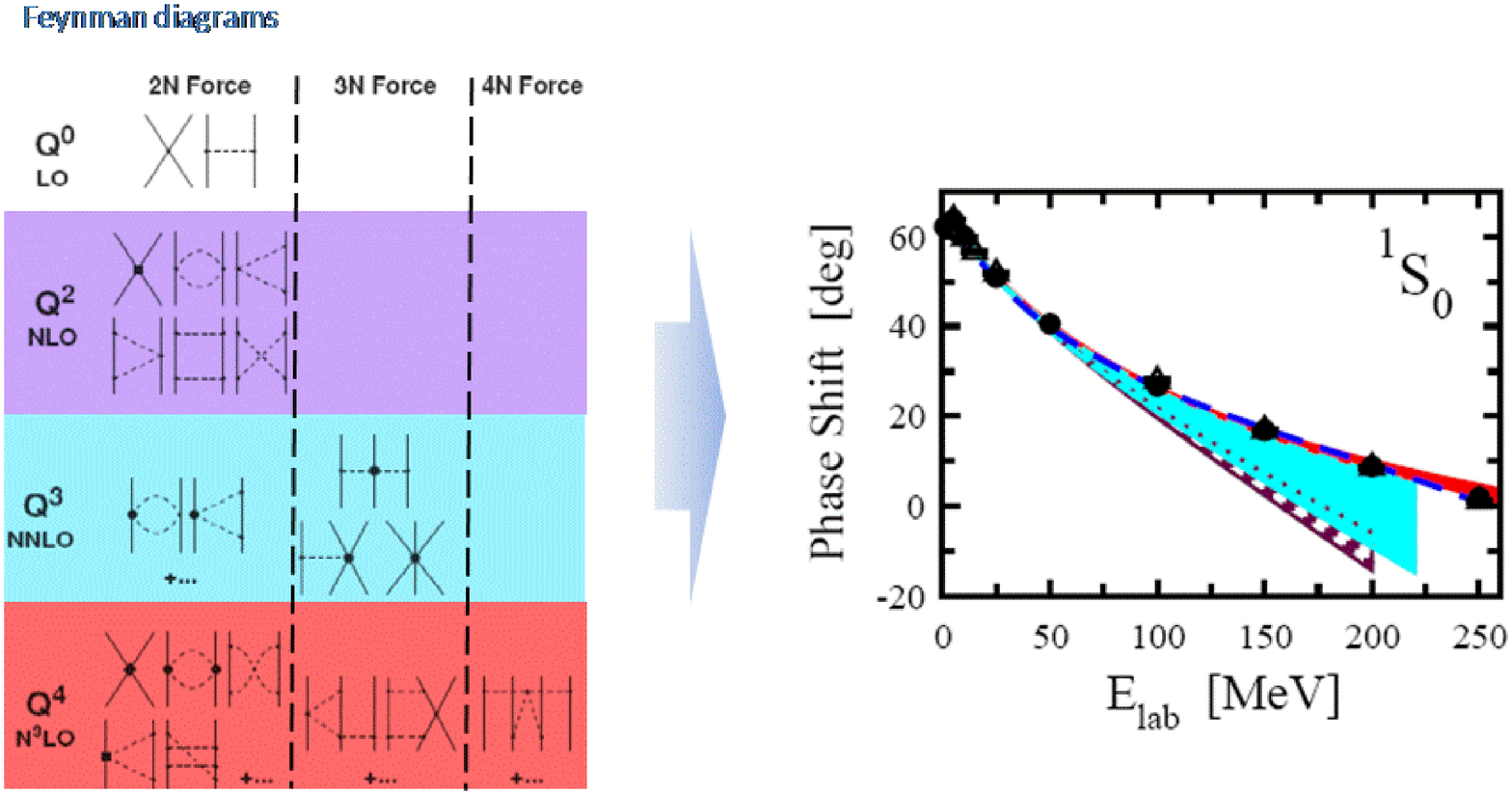}
\end{center}
\caption{(Color online) Left: Feynman Diagram representation of the different orders of EFT expansion. 
Right: Experimental phase-shift as a function of energy. The result of the calculation obtained at each 
order of the expansion are shown (from ref. \cite{Epe08}). Different colors correspond to different order shown with the same color
convention on the left.   
}
\label{fig:eft2}
\end{figure}

The EFT has been used starting from the QCD Lagrangian to provide a systematic and constructive framework 
for the strong interaction, i.e. 
\begin{eqnarray}
{\cal L}_{\rm QCD} \longrightarrow 
{\cal L}_{\rm EFT} = {\cal L}_{\pi\pi} + {\cal L}_{\pi N} + {\cal L}_{N N} + \cdots 
\end{eqnarray}
In that case, the typical energy range of interest is  $|\vec q| \sim M_\pi = 140$ MeV, while the cut-off 
parameter is of the order of the gap in energy that separates the $(u,d,s)$ quarks from the others, i.e. $1$ GeV.
An example of phase-shift analysis obtained using EFT up to $N^3$LO (next-to-next-to-next to leading order) is shown
in figure \ref{fig:eft2}. From this figure, it can be concluded that the calculated phase-shift converges to the 
experimental one as more and more orders in the expansion are included.   
Therefore, on opposite to conventional forces, EFT provides a systematic framework to obtain more and more accurate 
approximations for the strong nuclear interaction starting directly from QCD. In addition, it also gives consistent 
three-body, four-body interactions automatically (see figure \ref{fig:eft2} (right)).

An important aspects brought by EFT is the possibility to obtain interaction optimized for a specific 
energy scale. New interactions based on chiral EFT and discussed above still contain information on high energy that is not 
necessary for low energy nuclear physics. A second class of interactions, specifically dedicated 
to nuclei have been introduced, the so called "soft interaction" or "low momentum interaction.
This time, starting from either conventional interactions or interaction 
deduced from EFT, new interaction have been obtained using renormalization group (RG) techniques \cite{Keh06} while paying 
attention to keep the phase-shift at low energy unchanged when reducing further the cut-off compared to $\Lambda_{\rm QCD}$.        
\begin{figure}[htb]
\begin{center}
\includegraphics[width = 14.cm]{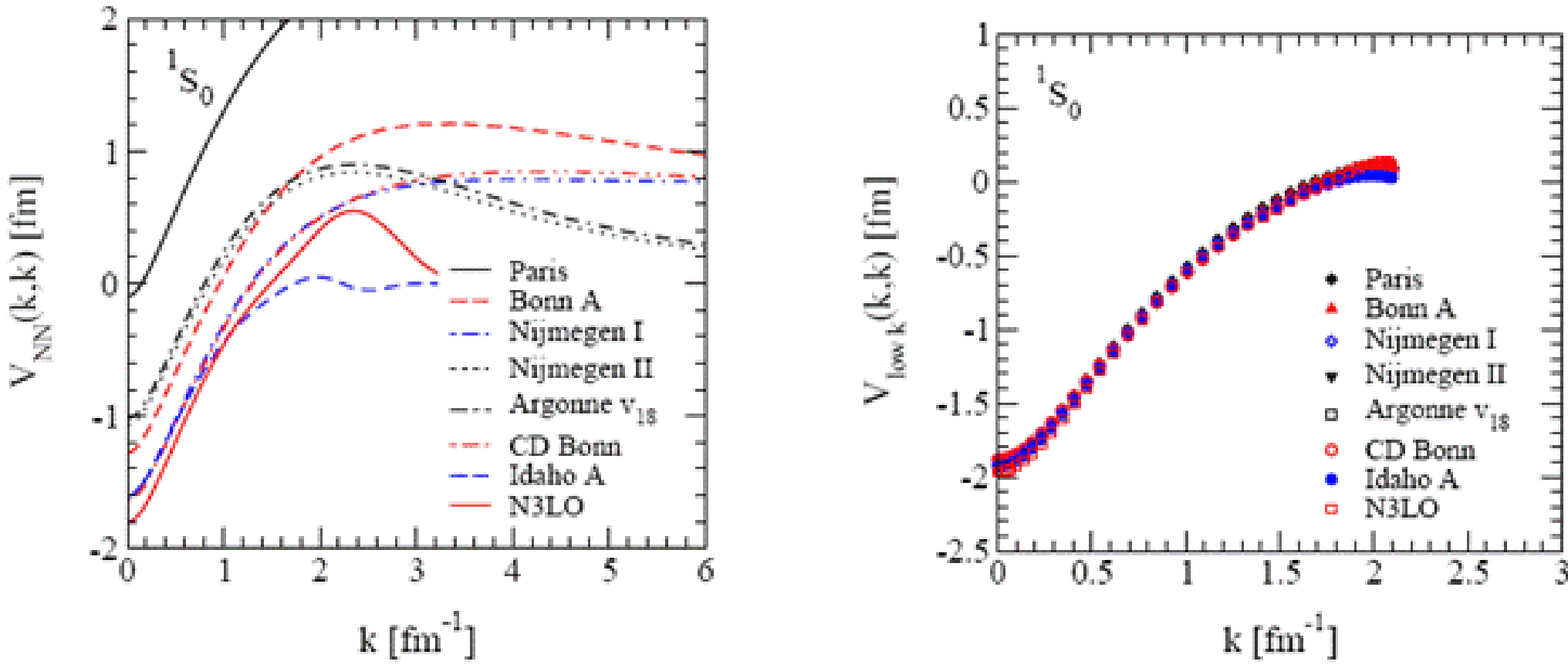}
\end{center}
\caption{(Color online) Left: Different conventional interaction as well as interaction obtained from EFT 
represented in momentum space. Right: New interactions deduced from the renormalization group procedure 
where the phase-shift has been constrained to remain unchanged \cite{Bog05}. 
In the latter case, all interactions that were 
originally completely 
different match with each other.}
\label{fig:vlowk1}
\end{figure}

The result that has been obtained is quite amazing. When the RG is systematically applied to different forces 
on the market, all interactions that were so different apparently lead to the same "universal" low momentum 
interaction. The conclusion from this study is that all forces were containing the same low energy information 
while the differences certainly stem from the high energy components which seems not to be constrained enough.
These new soft interactions, besides removing the hard-core difficulty, also provide a suitable answer 
to the "non-uniqueness" puzzle of conventional forces.      


\section{From nucleons to nuclei, hypernuclei and stars}

\begin{figure}[htb]
\begin{center}
\includegraphics[width = 14.cm]{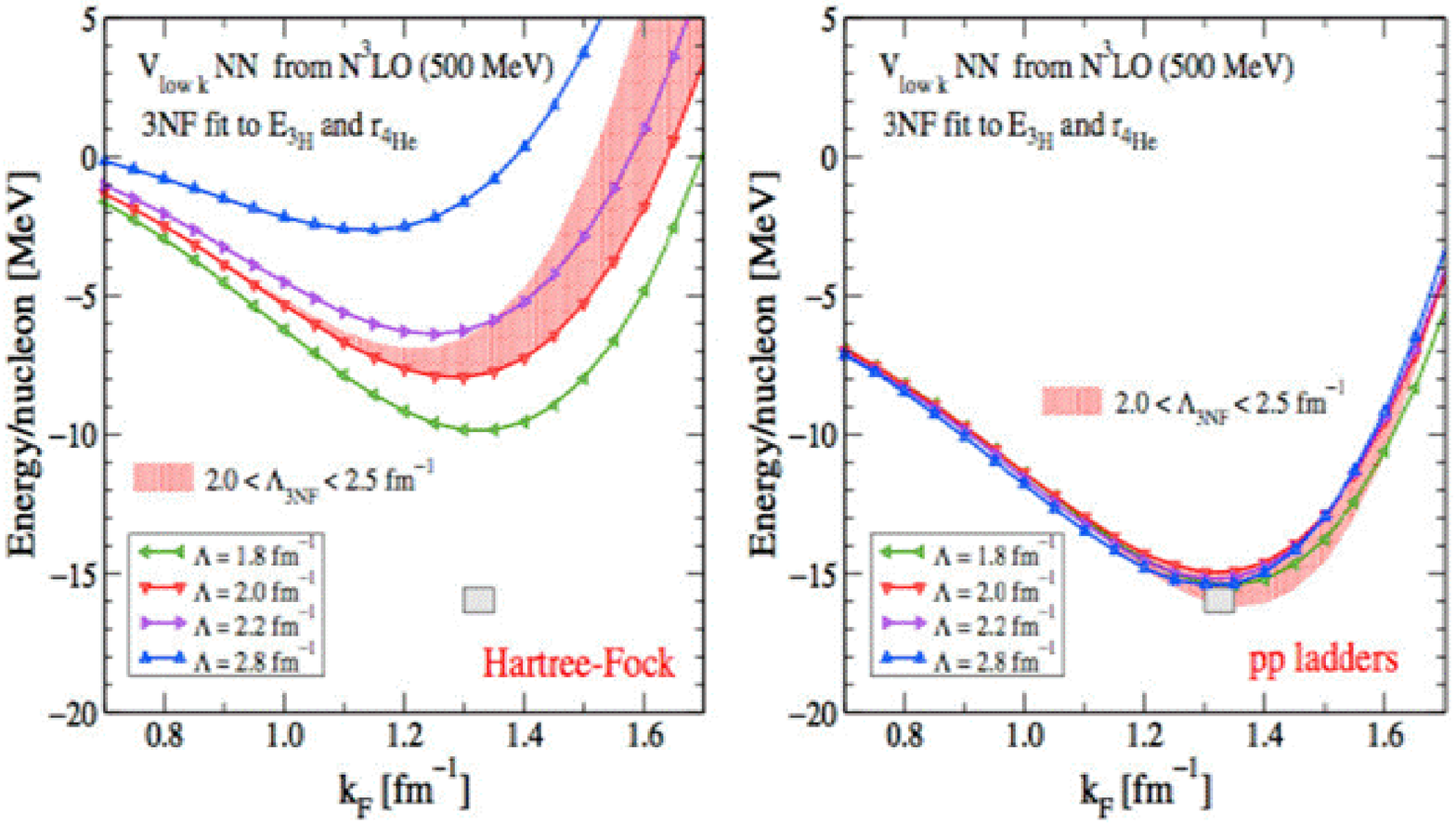}
\end{center}
\caption{(Color online) Energy per nucleon in symmetric nuclear matter as a function of the Fermi momentum calculated 
at the Hartree-Fock level (left), 
and including second order perturbation theory + particle-particle ladder contributions (right). 
The soft interaction used here is derived from the N$^3$LO nucleon-nucleon interaction. In both 
cases, different curves correspond to different cut-off parameters (for more detail see \cite{Bog09a})
Note that, a three-body force deduced from light nuclei properties is also used here (see discussion below).}
\label{fig:vlowk2}
\end{figure}

Conventional nuclear forces were suffering from three major drawbacks (i) lack of systematic constructive 
framework, (ii) non-uniqueness, (iii) difficulty to use due to the hard-core. As discussed above, recent advances 
in the theory of nuclear forces overcome these difficulties. New forces, especially soft ones, are expected to be 
much easier to use. This has been nicely illustrated in infinite nuclear matter. Up to know, the presence 
of a hard-core has made any attempt to use the strict independent particle (Hartree-Fock or Hartree-Fock Bogolyubov) 
approximation to the exact nuclear many-body problem useless. The recently proposed soft-core 
interactions, however, seem to make the nuclear
many-body problem perturbative \cite{Bog05,Bog09a}. The result of Hartree-Fock calculation for nuclear matter 
equation of state using the $V_{\rm lowk}$ interaction (left) and HF + second perturbation theory + particle-particle ladder
(right) are shown in figure \ref{fig:vlowk2} (taken from \cite{Bog09a}). Two important remarks should be made at this point: 
(i) while no reasonable results could be obtained using conventional forces with Hartree-Fock, a stable solution 
is obtained here already at this level of approximation (ii) Perturbation theory converges rather fast to the 
expected saturation point. Note that similar results can eventually be obtained with conventional forces but using 
much more involved many-body techniques.  

\subsection{Ab initio calculation}

\begin{figure}[htb]
\begin{center}
\includegraphics[width = 12.cm]{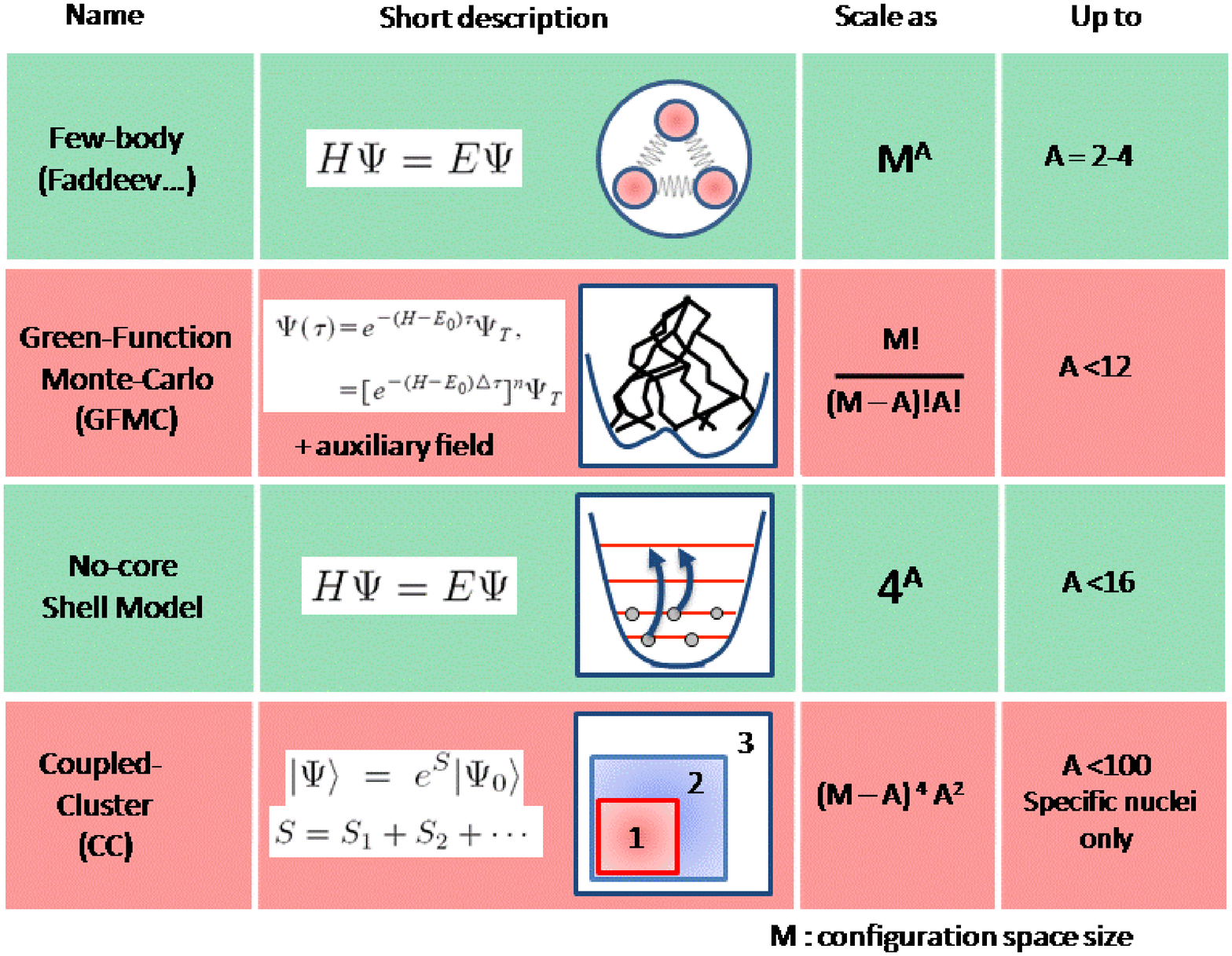}
\end{center}
\caption{(Color online) Schematic illustration of the Faddeev-Yakubowski, the  Green-Function Monte-Carlo (GFMC), 
the  No-Core Shell Model (NCSM) and the Coupled-Cluster (CC) theories (from top to bottom). 
In all cases, the basic equation that is used is shown as well as a logo summarizing the method, 
the computational cost in terms of system size and configuration space size as well as the actual 
(or estimated for the CC case) range of application are given in the last two rows respectively.  
}
\label{fig:abinitio_1}
\end{figure}

\begin{figure}[htb]
\begin{center}
\includegraphics[width = 8.cm]{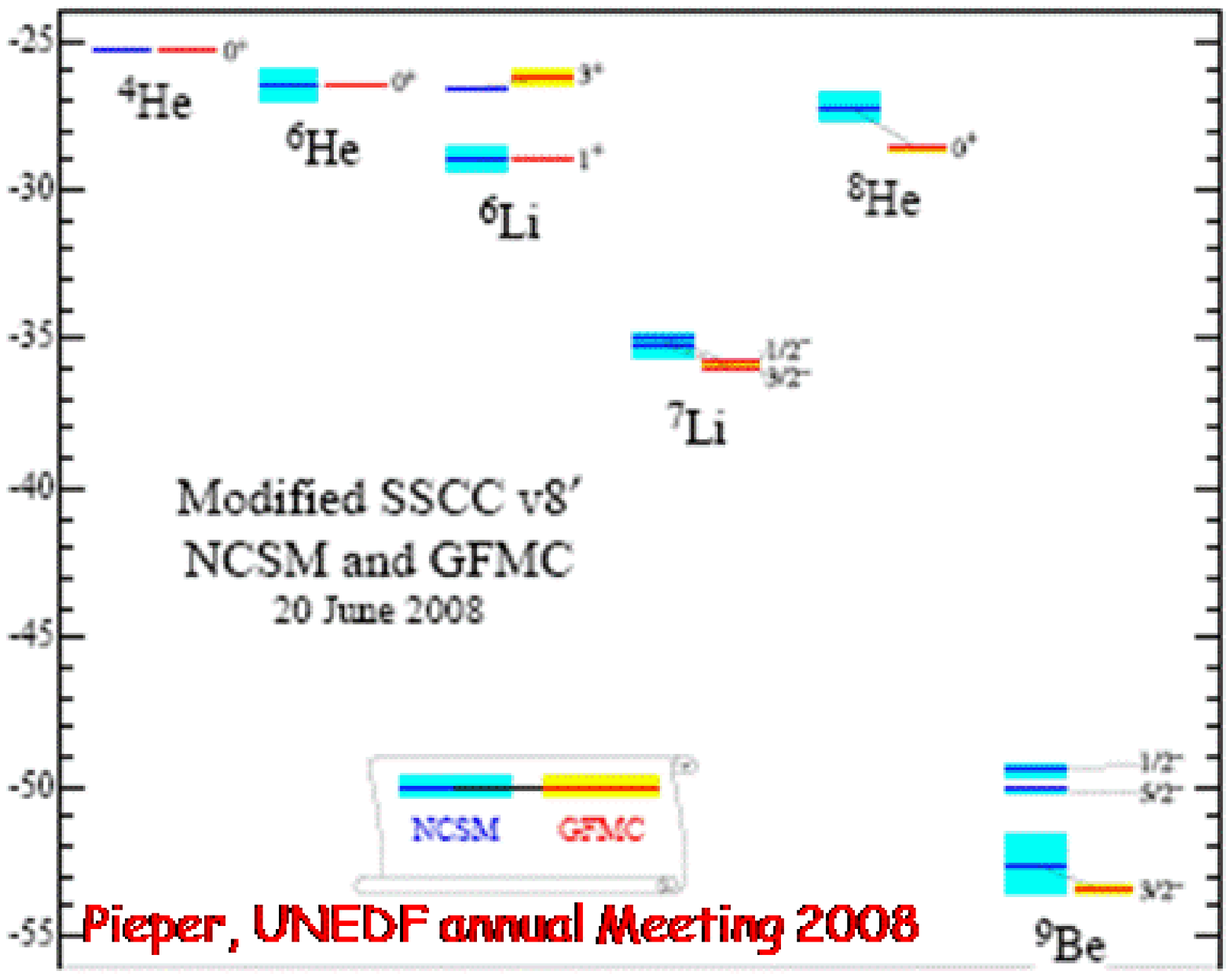}
\end{center}
\caption{(Color online) Comparison of results obtained with the NCSM and GFMC \cite{Pie04} for nuclei with $A \leq 9$.}
\label{fig:abinitio_3}
\end{figure}

Figure \ref{fig:vlowk2} illustrates that modern interactions, besides providing a direct connection with the underlying 
QCD Lagrangian, might also greatly simplify low energy nuclear physics calculations. In recent years, these forces have 
become a tool of choice for ab initio methods in light nuclei \cite{Bog09b}. 
With the increasing computational power, several theories have been recently developed to provide ab initio 
calculations of light nuclei properties, like for instance the  No-Core Shell Model (NCSM) \cite{Nav09}, 
the Green-Function Monte-Carlo (GFMC) \cite{Pie01,Pie07}, the Coupled-Cluster (CC) \cite{Kow04,Hag09}... A complete description 
of these theories is out of the scope of the present introduction. Nevertheless, figure \ref{fig:abinitio_1} illustrates 
schematically some of these theories.      
During the past decades important efforts have been made to make ab initio methods reliable. A benchmark of 
several methods on the $A=4$ \cite{Kam01} has shown that all methods are well under control and provide
the same results if the same interaction is used as an input. This is illustrated in figure \ref{fig:abinitio_3}   
where comparisons between NCSM and GFMC results are shown.     
Ab initio methods have reached a high degree of accuracy and can now serve as a benchmark for nuclear forces. 

\subsection{Light nuclei and three-body forces}

\begin{figure}[htb]
\begin{center}
\includegraphics[width = 10.cm]{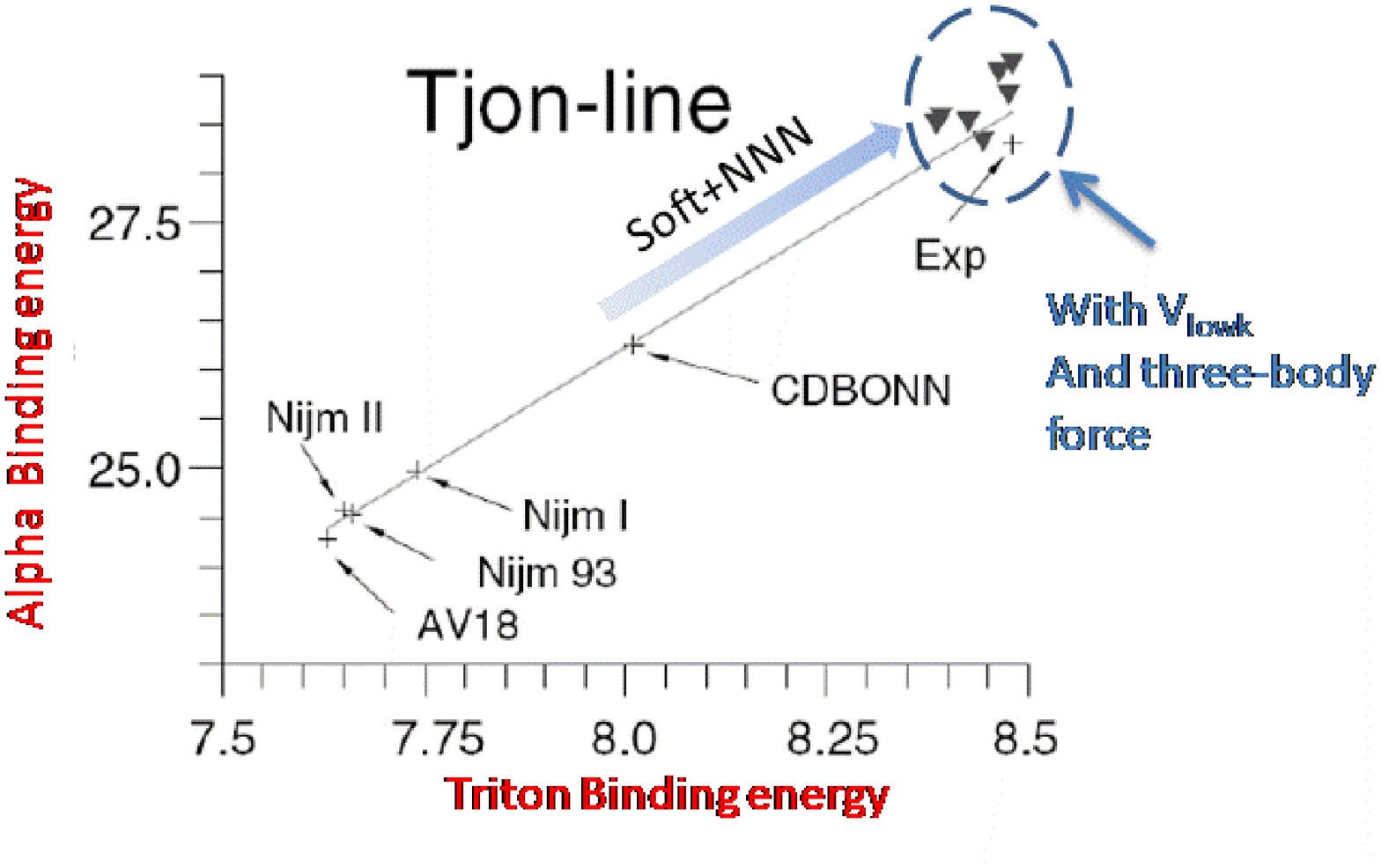}
\end{center}
\caption{(Color online) Experimental and theoretical binding energy of the alpha particle versus the same quantity 
for the triton \cite{Nog00}. Different theoretical points correspond to calculations with different nucleon-nucleon interaction.
Solid triangles correspond to calculations using soft interaction and 3-body nuclear force.}
\label{fig:3body_1}
\end{figure}

\begin{figure}[htb]
\begin{center}
\includegraphics[width = 14.cm]{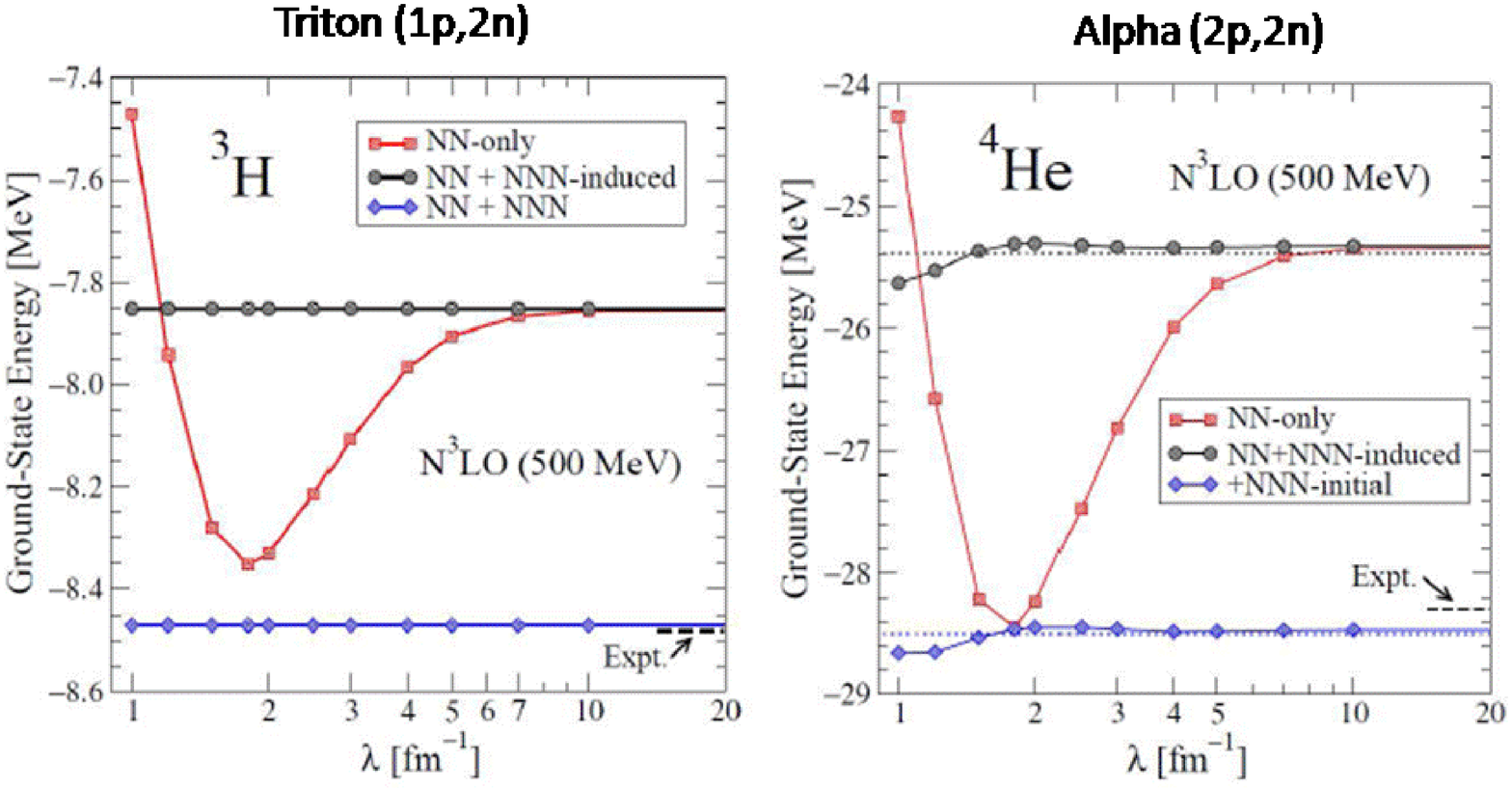}
\end{center}
\caption{(Color online) Example of recent application using modern forces described in section \ref{sec:modernforce}
combined with Similarity Renormalization Group (SRG) as a function of the SRG parameter $\lambda$ (for more details see 
\cite{Jur09}). Left: triton binding energy, Right: alpha binding energy. When only two body interaction are included, an 
induced three-body interaction is already present coming from two-by-two interaction. The filled circles and filled squares
correspond respectively to calculation using two-body interaction only with and without the induced 3-body 
force. Filled diamonds correspond to calculation when both the induced and bare three-body interaction are 
accounted for. Note that normally, for the alpha particle case, an induced 4-body interaction coming 
from the two and three-body force should a priori be included.     
}
\label{fig:3body_2}
\end{figure}

The actual trend of theoretical light nuclei studies is now to use a fully consistent bottom-up 
approach starting from QCD, deducing from it a bare nucleon-nucleon interaction dedicated to low energy 
nuclear physics and performing exact nuclear structure calculations. Recent discussions on the three-body 
force reflects the great interest of such a bottom-up approach. Applying ab initio methods to light nuclei with 
two-body forces only cannot reproduce experimental observation. This is shown in Figure \ref{fig:3body_1} 
presenting the experimental or theoretical binding energy of the alpha particle versus the same quantity 
for the triton \cite{Nog00}. As seen in this figure, without 3-body interaction, both alpha and triton binding 
energy are underestimated while if it is included, the experimental data can be reproduced.    
It is worth mentioning that conventional forces discussed in section \ref{sec:conventional}, were already not 
able to reproduce light particle binding energy. However, due to the flexibility of such forces it is difficult 
to have a definite answer about the three nucleon interaction. In particular, one may have though that a slight 
modification of the two nucleon force can eventually improve the agreement with experiments. The introduction 
of modern forces based on EFT leaves less room for adjustment showing that a 3-nucleon force is definitely 
missing. An illustration of most recent ab initio calculation starting from chiral perturbation theory and EFT 
up to N$^3$LO is shown in figure \ref{fig:3body_2}. The agreement of the full 3-body calculation with the 
experiment show the amazing degree of accuracy of modern forces. From these different curves, one may also 
estimate the importance of different forces components contribution. In table \ref{tab:percentforce}, the percentage of contribution of the two and three-body forces
on the binding energy of triton and alpha particle are estimated. Using the difference between the calculated
alpha particle binding energy and the experimental one, an upper estimate of a possible four-body force is also given.    
\begin{center}
\begin{table}[bthp]
\begin{tabular} {l|cc}
\hspace*{4.cm} &  triton & alpha \\
\hline 
2-body force &  \hspace*{1.cm} 92 \% \hspace*{1.cm}  & \hspace*{1.cm}  89 \% \hspace*{1.cm}  \\
3-body force &  7 \%  & 10 \% \\
4-body force & - & $<$ 1 \% \\
\hline
\end{tabular}
\caption{\label{tab:percentforce}
Estimate of the percentage of contribution of $2$-, $3$- and $4$-body forces in the binding 
energy of triton and alpha particles. Percentage here are obtained by using the formula 
$(E_{\rm Cal} - E_{\rm exp})/E_{\rm exp}$, where $E_{\rm Cal}$ correspond to the calculated 
binding energy at different levels of approximation.}
\end{table}
\end{center}

\subsection{Hypernuclei}

A similar bottom-up strategy is also being followed to provide 
ab initio estimates for Hypernuclei starting from QCD ingredients. As discussed earlier, hyperons 
correspond to nucleon where one of the $u$ or $d$ quarks is replaced by a $s$
quarks. Therefore, all the discussion made above should be extended to the $(u,d,s)$
quark space. Fortunately, the $s$ quark mass (see (\ref{eq:massquarks})) is also much below the QCD gap (1 GeV)
and one can take advantage of the EFT machinery to provide hyperon-nucleon 
or hyperon-hyperon interaction \cite{Pol06}. 

\begin{figure}[htb]
\begin{center}
\includegraphics[width = 14.cm]{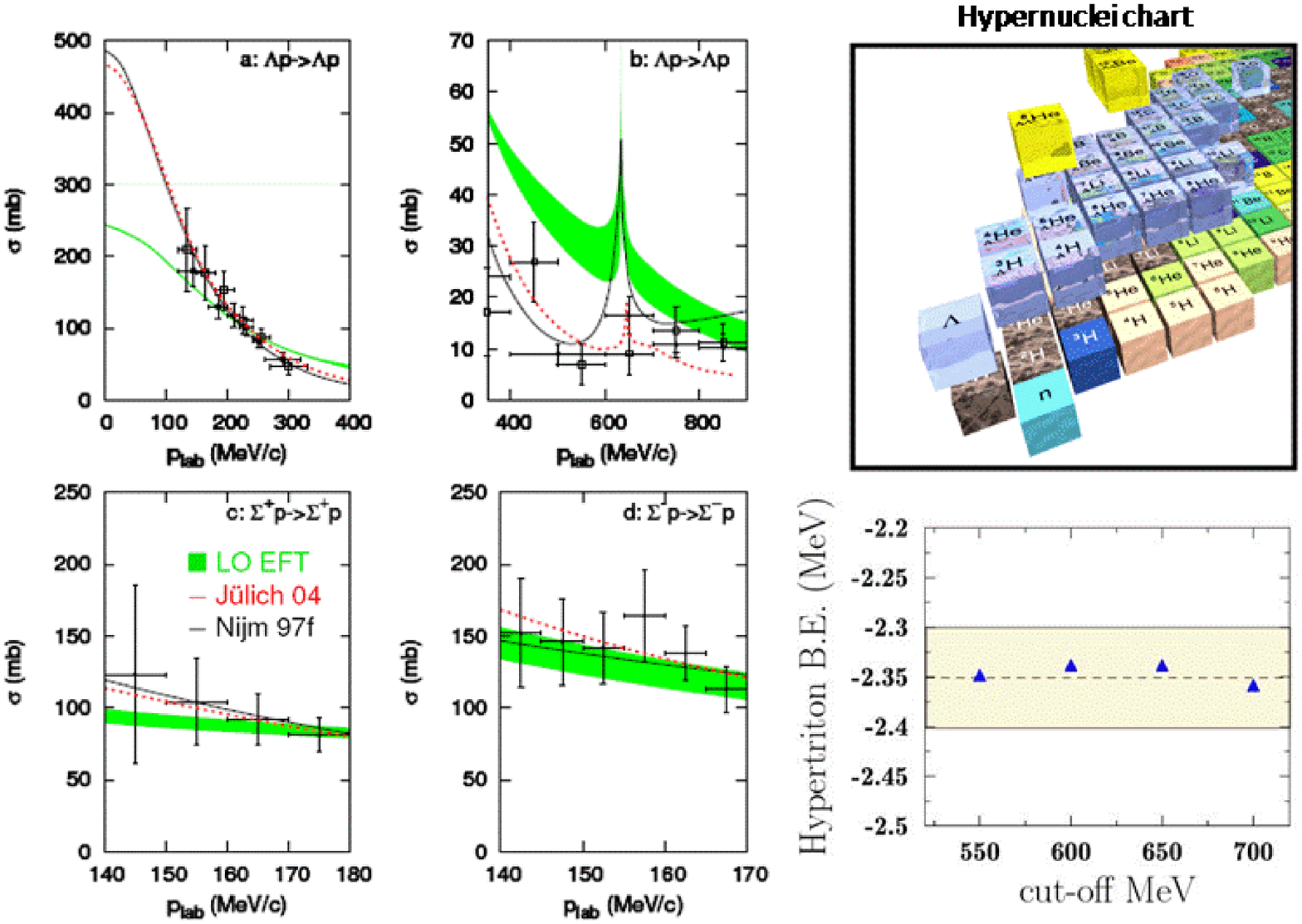}
\end{center}
\caption{(Color online) Left: Comparison of measured hyperon-nucleon cross-section with calculated 
ones estimated using EFT up to NLO (taken from \cite{Pol06}). Right-top: the hypernuclei chart. Right-bottom: Example of result of
ab initio calculation for hypertriton. The dashed area corresponds to the experimental value with errorbars, while 
filled triangle are calculated values as a function of the cut-off parameter.}
\label{fig:hypernuclei}
\end{figure}

An illustration of calculated hyperon-nucleon cross section obtained from chiral EFT is presented in figure 
\ref{fig:hypernuclei} and compared with experimental cross sections. These cross-sections have large errorbars 
due to the limited number of data set (around 30-40) existing for hyperons compared to the nucleon-nucleon 
case (between 2000 and 3000).  Results obtained in ref. \cite{Pol06} show that at NLO, the theoretical 
prediction are already in rather good agreement with experimental data. In figure \ref{fig:hypernuclei}, 
an example of ab initio
calculation for the hypertriton is also shown (lower right panel) giving a binding energy within the experimental 
errorbars.     

\subsection{New Trends in the description of medium and heavy nuclei}

\begin{figure}[htb]
\begin{center}
\includegraphics[width = 10.cm]{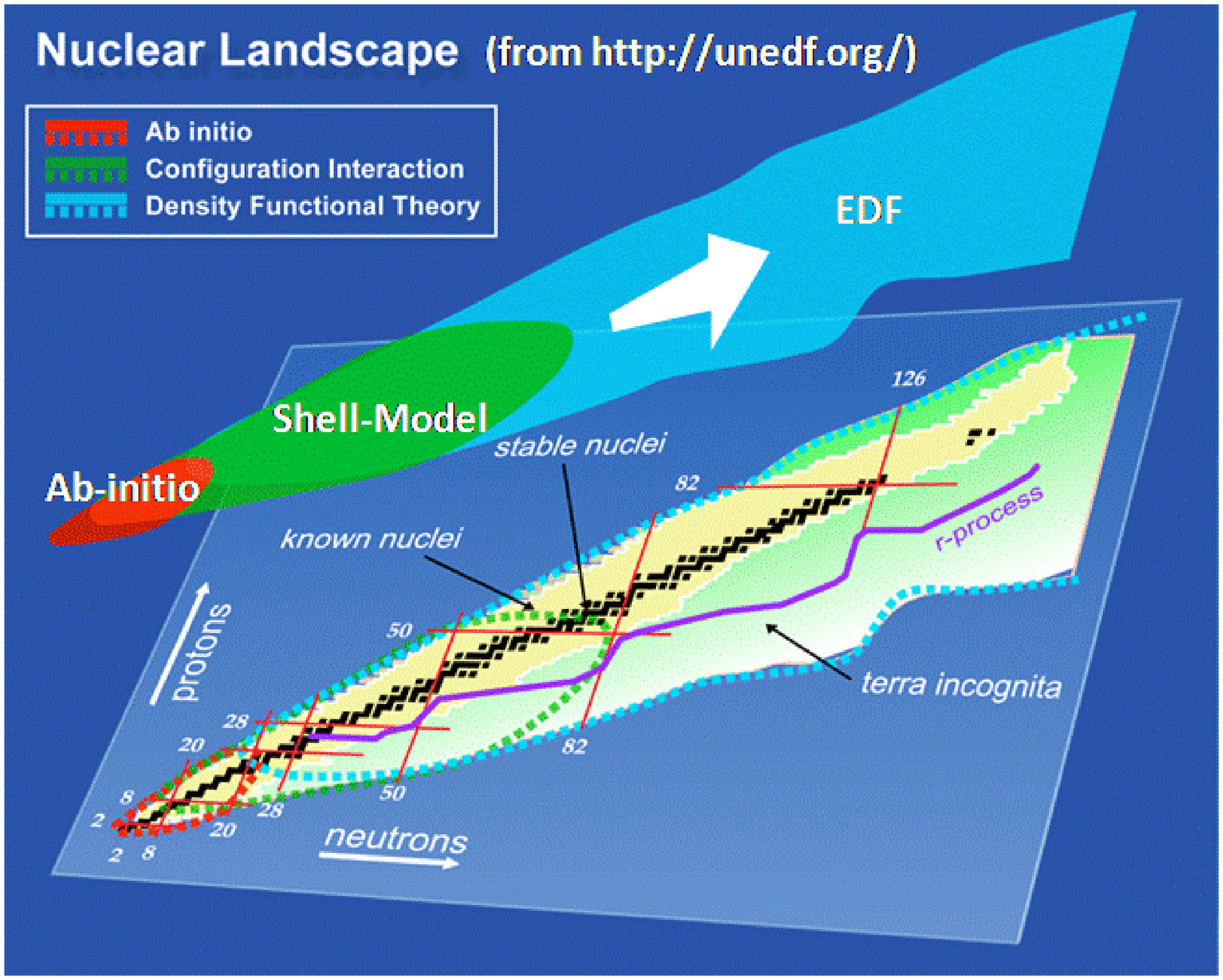}
\end{center}
\caption{(Color online) Summary of the different microscopic approaches dedicated to the nuclear many-body problem 
(adapted from the UNEDF (Universal Nuclear EDF) project website \cite{Ber07,Ber07a}.). 
While ab-initio methods are restricted to light nuclei, nuclear shell model is expected to describe nuclear structure 
in medium nuclei ($A< 100$) in the near future. Meanwhile, the EDF theory is the only theory able to treat 
nuclear system over the whole nuclear 
chart.}
\label{fig:unedf}
\end{figure}

Due to their numerical complexity, ab initio methods are limited to rather light nuclei. One of the great challenge of todays 
nuclear physics is to provide a microscopic theory able to describe nuclei over the whole nuclear chart. Considering first nuclear 
structure problem, two theoretical frameworks have become standard tools for low energy nuclear physicists, namely the 
Shell Model \cite{Cau05} and the Energy Density Functional theory \cite{Ben03,Sto07}. Figure \ref{fig:unedf} summarizes 
the range of applicability of most popular nuclear theories. Up to now, nuclear models contain a large number of 
phenomenological inputs (monopole interaction, residual interaction, components of the functionals ...) largely 
adjusted to reproduce experiments. As a result, the link with the underlying bare nucleon-nucleon interaction is often 
lost. In the mean time, recent studies have pointed out the lack of predicting power for not yet experimentally 
explored area of the nuclear chart. Two strategies exist to overcome this difficulty: (i) either, new experiments specifically 
dedicated to unexplored region are performed in order to better constrain phenomenological inputs of nuclear models (ii) or 
new approaches, following the bottom-up spirit, where the connection with the bare interaction and QCD is kept, are developed.     

In this section, recent examples demonstrating the new perspective offered by 
modern nuclear forces and using the second strategy are illustrated. 

\subsubsection{Nuclear Shell Model}

\begin{figure}[htb]
\begin{center}
\includegraphics[width = 14.cm]{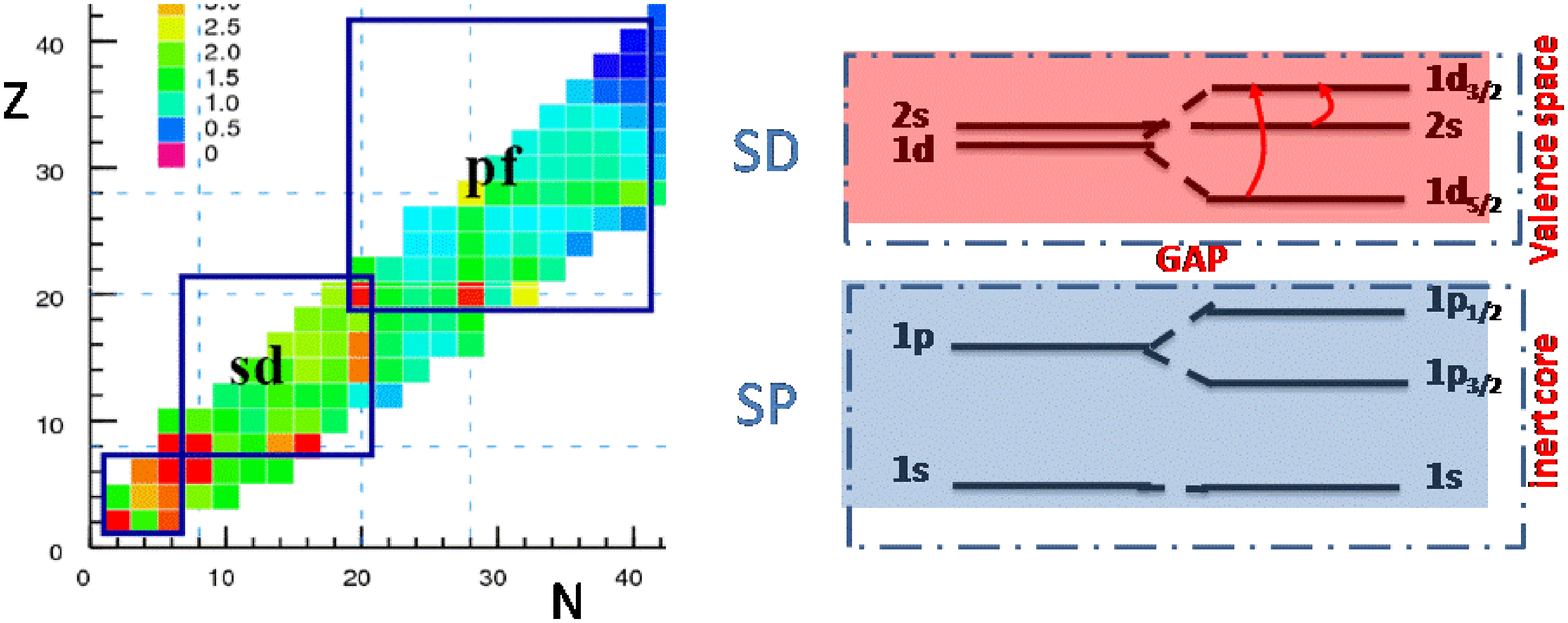}
\end{center}
\caption{(Color online)Left: Illustration of the different single-particle space generally used in the shell model 
(Adapted from \cite{Bro07}). Right: example of inert core and valence space used to describe $sd$-shell nuclei.}
\label{fig:shell}
\end{figure}   

\begin{figure}[htb]
\begin{center}
\includegraphics[width = 14.cm]{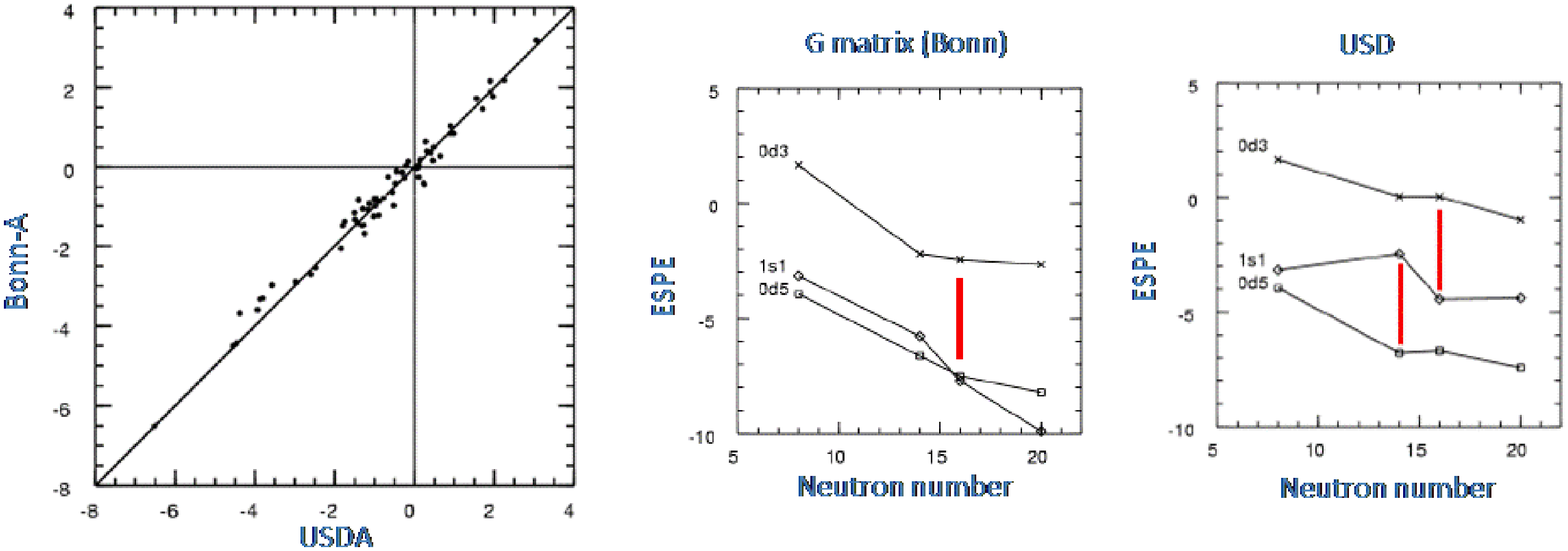}
\end{center}
\caption{(Color online) Left: Correlation between the residual interaction components obtained through the G-matrix technique 
starting from the Bonn interaction and the corresponding components after re-adjusting to better fit the experiments 
(Adapted from \cite{Bro01,Bro07}). Right: Effective Single Particle Energy (ESPE) deduced in both cases for the oxygen 
isotopic chain.}
\label{fig:brown_espe}
\end{figure}

\paragraph{\bf Basic discussion on Shell Model:} Here, the Shell Model approach is presented in a very schematic way and readers interested 
in all the subtle aspects of this approach can refer to the excellent recent review articles \cite{Bro01,Ots01,Cau05}.
 Standard Shell model starts from a single particle basis  
and construct from it Many-Body waves function written as 
\begin{eqnarray}
\Psi = \Phi_{[0p0h]} + \Phi_{[1p1h]} + \Phi_{[2p2h]} + \cdots  
\end{eqnarray}
where $\Phi_{[0p0h]}$ denotes the independent particle state, while $\Phi_{[1p1h]}$, $\Phi_{[2p2h]}$...
denotes the mixing of different $1$ particle-$1$hole,  $2$ particle-$2$ hole ...excitations. Many-Body 
states are obtained by direct diagonalization of an effective two-body Hamiltonian, that is schematically 
written as:
\begin{eqnarray}
H &=& {\rm SPME} (\varepsilon_i) + {\rm TBME} (G) 
\end{eqnarray}   
where SPME and TBME stands respectively for Single-Particle and Two-Body Matrix Elements, denoted respectively by 
$\varepsilon_i$ and $G_{ijkl}$. Due to the numerical complexity, standard Shell Model generally separates 
the single-particle space into an inert core and valence states that participate to correlations 
(see illustration \ref{fig:shell}). Accounting for the interaction symmetries, within a finite space of single-particle, 
we are left with a finite number of input parameters $\{\varepsilon_i, G_{ijkl} \}$. In conventional 
Shell Model single particle energies are 
generally directly adjusted to reproduce experiments. $G_{ijkl}$ matrix elements  can a priori be deduced from 
the G-matrix theory starting from a bare interaction that is dressed to properly account for medium effects. However,     
direct use of G matrix components do not lead to satisfactory results. For this reason, $G$ components are in 
practice slightly readjusted to better reproduce a set of spectroscopic properties in a number of 
nuclei relevant for the selected single-particle space. For instance, in the $sd$-shell, this leads to 
66 parameters that have been adjusted on more than $600$ levels to get a deviation lower than 200 keV.   
In figure \ref{fig:brown_espe} (left), the correlation between 
the G matrix components directly deduced from a conventional bare interaction and their values after re-adjustment 
is presented \cite{Bro01,Bro07}. Corresponding effective single-particle energies (ESPE) are also shown in this figure
for the oxygen isotopic chain.  
Two important remarks could be drawn from Fig. \ref{fig:brown_espe}: (i) there is a strong correlation between 
the initial value of the residual interaction and the final one showing that the re-adjustment is small. (ii) Even 
a small modification of the matrix elements leads to a rather large effect on ESPE and is for instance crucial to understand 
the appearance of new gaps.

The Shell Model is certainly the most precise tool available to the nuclear structure community. Insight into 
experimental observation can generally hardly be made without this approach. The understanding of new gaps related to
the discovery of new island of stability strongly depends on the SM interpretation. Conjointly, one of the goal 
of studying isotopic and/or isotonic nuclei chains is to be able to uncover properties of the nuclear force
in the nuclear medium. For instance, change of ESPE has led to important discussion on the spin-orbit coupling 
or more recently on the tensor part of the interaction \cite{Ots07,Sor08}.         
However, at the sight of the strong dependence of the ESPE on the residual interaction (see figure \ref{fig:brown_espe})
as well as the phenomenology introduced during the re-adjustment of the G matrix, 
one may worry about the possibility to draw
precise conclusion on specific components of the interaction.  
\\

\begin{figure}[htb]
\begin{center}
\includegraphics[width = 14.cm]{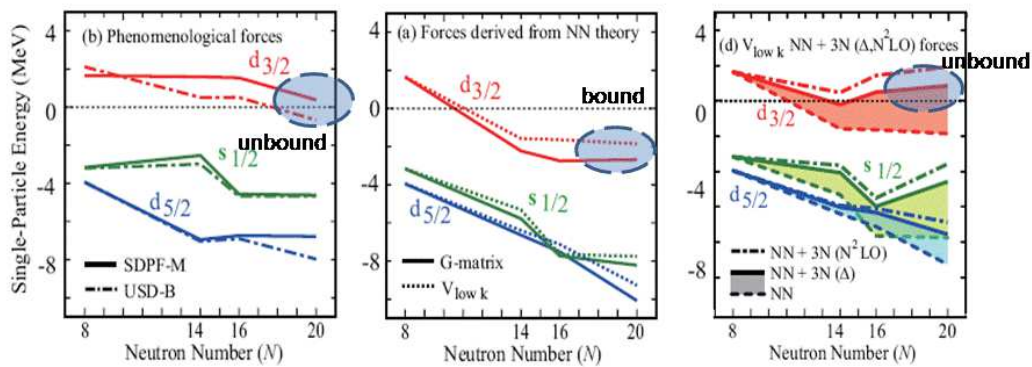}
\end{center}
\caption{
(Color online) ESPE for the oxygen isotopic chain obtained in phenomenological shell model (Left), 
using directly the G matrix components deduced from either conventional or $V_{\rm lowk}$ two-body interaction (Middle)
and the same G matrix components with the $\Delta$
N$^2$LO part of the three-body interaction (Right) (adapted from \cite{ots09}). 
}
\label{fig:otsuka}
\end{figure}

\paragraph{ \bf New trends in the nuclear shell model:}
The fact that a re-adjustment is needed to understand experiments reflects
that part of the physics relevant for nuclear structure is missing in the bare 
two-body interaction input given to the Shell Model. 
According to recent discussions, the absence of three-body interaction in conventional Shell Model 
appears as a good candidate. Taking advantage of the EFT constructive framework to provide three-body 
interaction (figure \ref{fig:eft2}), the Shell Model theory has been recently applied using soft interaction and part 
of the three-body interaction (the component corresponding to a $\Delta$ excitation appearing at the N$^2$LO order).
Corresponding ESPE are given in figure \ref{fig:otsuka} and compared to ESPE obtained in phenomenological 
Shell-Model and/or with ESPE obtained when the three-body component is omitted (adapted from \cite{ots09}).       
Although the application presented in figure \ref{fig:otsuka} is still neglecting many components of the three-body
interaction that could be deduced from EFT, it paves the way for future application of the nuclear Shell-Model. It 
also brings many interesting aspects: 
\begin{itemize}
\item The expertize acquired in the Phenomenological Shell-Model is crucial. Indeed, ESPE energies presented in left side 
of figure \ref{fig:otsuka} corresponds to the best ESPE value optimized to reproduce actual nuclear data. It therefore will 
serve as a reference for future calculation directly based on modern forces without re-adjustment.
\item Not surprisingly, new soft interaction leads sensibly to the same ESPE as conventional forces. Indeed, the same 
information on low energy nuclear physics is contained in both interactions. The only expected difference is to simplify a priori 
the "dressing" of the force due to the absence of a hard-core. 
\item Although the effect of the three-body interaction is expected to be much smaller than the two-body contribution, 
it significantly affects the ESPE in nuclei.  
\item Last, the influence of the three-body interaction and the fact that (at least for the unbound state) it becomes 
closer to the phenomenological prescription tends to precise the re-adjustment role. Indeed, it is reasonable 
to conclude that the slight re-adjustment was a way to include three-body and higher-order effects that were not accounted 
for in the input bare two-body interaction at the first place.  
\end{itemize}
     
The last remark also illustrates one of the drawback of phenomenological Shell-Model. Indeed by fitting directly 
experiments, adjusted residual interaction, even if they are very close to those obtained after the G-matrix application, mixes
several effects. It is then very difficult to disentangle these effects and easy to misinterpret shell evolution. This is the 
challenge of future Shell Model application based on new forces and used without re-adjustment to be able 
to reproduce experiments and get better physical insight in nuclear structure study.           

\subsubsection{Energy Density Functional theory}

\begin{figure}[htb]
\begin{center}
\includegraphics[width = 14.cm]{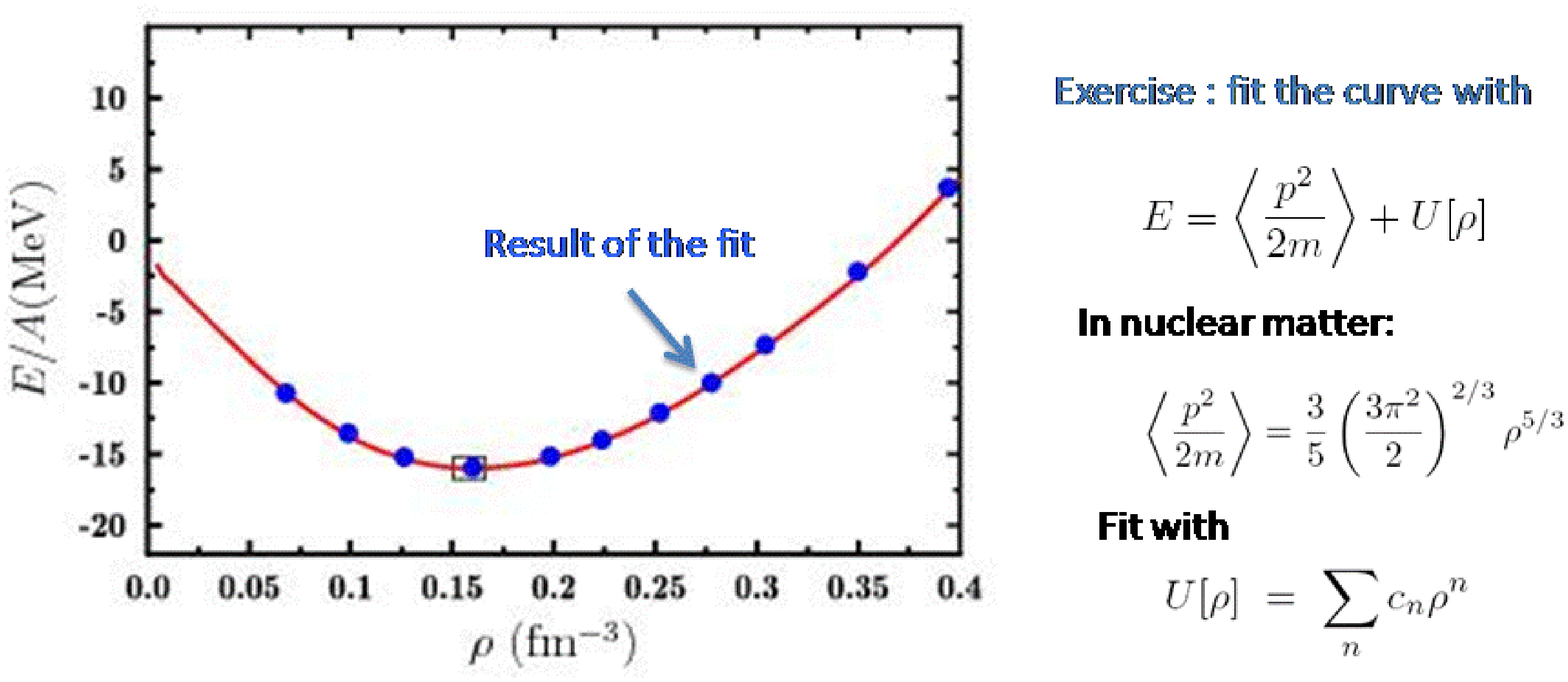}
\end{center}
\caption{(Color online) Consider a reference curve that gives the energy of symmetric nuclear matter as a function 
of its density (here, the red curve represent the ab-initio calculation performed in ref. \cite{Fri81}). The goal 
is to fit the curve by energy written as a functional of the density of an independent particle state. 
The first step is to write the functional as a sum of the kinetic and potential energy term. For independent particle 
system, the kinetic term directly express as a functional of $\rho$. As an illustration, it is assumed here than the potential 
part $U[\rho]$ simply writes as a polynomial of the density. It is straightforward to show that a fit (blue circles) 
with a fifth order polynomial gives a perfect reproduction of the curve. }
\label{fig:edfbasic}
\end{figure}   

\paragraph{\bf Basic discussion on Energy Density Functional:}
The nuclear Energy Density Functional (EDF) is very close in spirit to the Density Functional Theory  
(DFT) introduced in electronic systems \cite{Par89,Dre90,Koc01,Fio03}. The great advantage of DFT (and EDF)
is to establish a mapping between the original, most often intractable, 
many-body problem of interacting particles and a functional theory that can be solved using 
an independent particle method. In its simple form, where an auxiliary determinant state 
is used to construct the one-body density $\rho$, this mapping can be presented as:
\begin{eqnarray}    
E = \frac{\left\langle \Psi |H | \Psi \right\rangle}{\left\langle \Psi | \Psi \right\rangle} ~ \Longleftrightarrow  ~ 
{\cal E}_{\rm EDF} [ \rho ] ~\Longleftarrow ~ \rho_{ij} &=& \langle \Phi |a^\dagger_j a_i| \Phi \rangle 
~ \Longleftarrow  ~ |  \Phi \rangle =  \prod a^\dagger_i | - \rangle ~.
\end{eqnarray} 
Historically, the introduction of functional theory was not motivated by firm existence theorem like 
the DFT in condensed matter but introduced more empirically having in mind the quite simple evolution of the 
ground state energies and universal behavior 
of the nuclei densities along the nuclear chart \cite{Vau72,Neg72}. Although most nuclear physics EDFs start
from an effective interaction (Skyrme like \cite{Vau72} or Gogny like \cite{Dec80}), a more direct and simple 
illustration of EDF 
construction is given in figure \ref{fig:edfbasic} to illustrate the EDF spirit. When effective interaction 
are used, the parameters of the interaction are directly adjusted to reproduce properties of finite and infinite 
nuclear systems. This strategy is called phenomenological or empirical EDF hereafter.

\begin{figure}[tbhp]
\includegraphics[width=16.cm,clip]{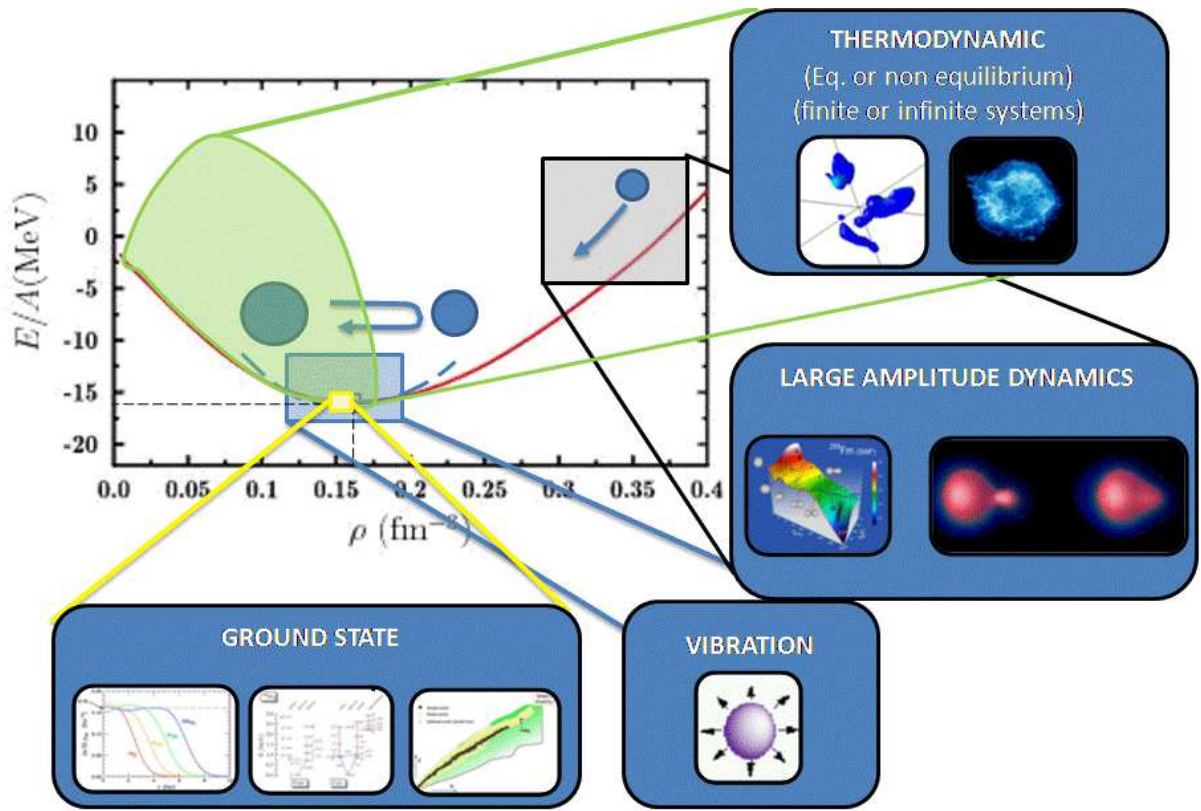}
\caption{ {\bf Range of application of the nuclear EDF}: (Color online) The red curve represents the equation of state in symmetric nuclear matter as a function 
of the system density. In the EDF theory, the energy is directly parametrized as a functional of the density. 
Nuclear structure study focus on or around the minimum. 
EDF is used extensively to get information on the ground state (yellow box) as well as excited state (see discussion in the text).
It could also be used when the system is slightly shifted from the minimum leading to the onset of small amplitude vibrations
(blue box). Time-Dependent version of the EDF also provides a description of nuclear dynamics like fusion or fission when the system 
is far away from the minimum (large amplitude collective motion [LACM]). Finally, EDF can also be extended to treat systems at finite 
temperature or entropy and provide information on the full phase-diagram and associated phase-transition.     
} \label{fig:edf0}
\end{figure}

The introduction of EDF in nuclear system in the early 70's was a major breakthrough.
Today, the EDF is not restricted to ground state properties but is expected to provide 
a unified microscopic framework  able 
to address the diversity of phenomena taking place in nuclei from nuclear structure to nuclear 
reactions or nuclear astrophysics: nuclear spectroscopy, small and large amplitude dynamics, equilibrium and 
non-equilibrium thermodynamics (see illustration in figure 
\ref{fig:edf0})...


\begin{figure}[htb]
\begin{center}
\includegraphics[width = 16.cm]{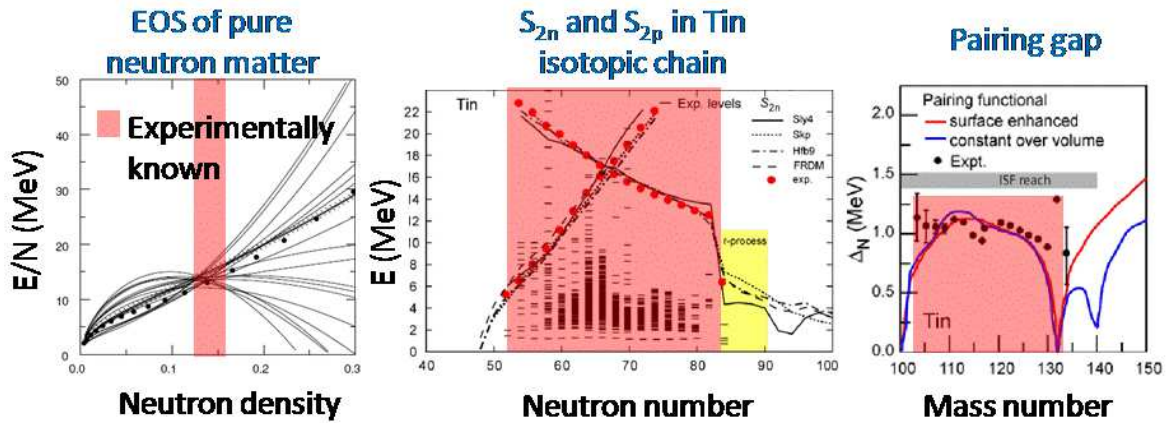}
\end{center}
\caption{(Color online) Left: Energy of the neutron matter as a function of density (taken from \cite{Bro00}), 
Middle: Two neutron and two proton 
separation energies along the Tin isotopic chain \cite{Msu07}, Right: Pairing gap in the Tin isotopic chain \cite{Msu07}. 
In all cases, filled 
circles correspond to reference values obtained either by ab initio methods (left) or directly provided by experiments 
(middle and right). Well constrained regions are indicated by pink shaded area. In all cases, different curves correspond 
to different phenomenological approaches most often using different sets of Skyrme interaction parameters values.     
}
\label{fig:edfdrawback}
\end{figure}

Most recent EDF are able to reproduce ground state energy with a deviation around $500-600$ keV over the whole nuclear chart 
which is already very good in view of the limited parameters number (less than 20) used to design the functional. 
Still, phenomenological EDF face the difficulty of large dispersion where no data exist (see examples in figure 
\ref{fig:edfdrawback}) 
showing the lack of constraints in unexplored regions. Figure \ref{fig:edfdrawback} clearly shows that the same EDF 
theory with different parameters values have been properly adjusted to provide a fair description of existing 
observations. However, they significantly differ from each others far from stability.     
\\

\paragraph{\bf New trends in EDF:}
For the future of EDF theory and to increase its predictive power, it is highly desirable to reduce
this uncertainty. Two options are now being developed to reach this goal: either add new observations in unknown region 
by studying nuclear systems under extreme conditions (low/high density, large isospin...) or improve the theory to 
make it less phenomenological. 
   
The first option goes with the use of new radioactive beam facilities that will open new perspectives for extracting 
specific contributions to nuclear properties. One example is the quest of symmetry energy extraction at various 
density. Large effort is actually devoted to this issue for 
instance using Heavy-Ion reactions \cite{Bar05,Dit07,Li08} and/or 
Nuclear Astrophysics inputs \cite{Lat00,Ste05,Lat07b,Xu09}.
          
Here, I will devote the rest of this section to the second option where the ultimate goal is to 
provide new EDF. Illustrations given in figure \ref{fig:edfdrawback} point out the limitation
of conventional EDF based on direct fit of experimental data. In addition, similarly to the 
conventional Shell-Model case, once a fitting procedure is used, it is very difficult, not to say impossible 
to make connection between the EDF parameters values and the bare nuclear force. For instance, in the 
original formulation of EDF based on Skyrme force, a three-body contact interaction is introduced. This 
interaction has nothing to do with the bare 3N interaction and is just a practical way to obtain            
a third order density dependence of the energy functional. Finally, recent pathologies 
observed in EDF when combined with configuration mixing (see for instance \cite{Lac09b,Ben09,Dug09})  
illustrate the lack of precise theoretical foundation for such a theory. As mentioned previously, on opposite 
to DFT, nuclear EDF has been introduced in a rather empirical way. This approach is being now revisited to put it on 
a firm theoretical ground. For instance, EDF is based on the existence of a functional of the intrinsic density for which, 
up to recently, an existence theorem was missing \cite{Eng07,Gir08,Mes09a}. EDF practitioners use and abuse of symmetry 
breaking (particle number, spatial and rotational invariance...) to be able to grasp static and dynamical correlations 
in nuclei with relatively simple functional \cite{Ben03}. The possibility to use the symmetry breaking and the associated 
symmetry restoration within a functional framework is far from being trivial and is a key issue that has to be clarified
in the near future \cite{Dug10}.  

\begin{figure}[htb]
\begin{center}
\includegraphics[width = 12.cm]{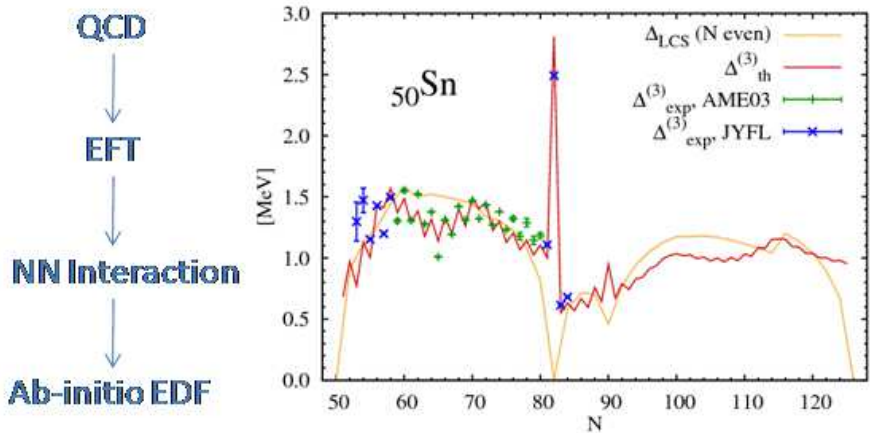}
\end{center}
\caption{(Color online) Left: Schematic representation of the full bottom-up approach connecting QCD to ab initio 
EDF. Right: Pairing gap in Tin isotopes obtained by directly plugging the bare $V_{\rm lowk}$ interaction (and Coulomb interaction for protons) 
into the pairing channel while keeping the mean-field channel unchanged. The final result denoted by $\Delta^{(3)}_{\rm th}$ 
compares perfectly with the experimental gap $\Delta^{(3)}_{\rm exp}$ \cite{Dug09b}. }
\label{fig:pairinggap}
\end{figure}

Assuming that above aspects are clarified, the second challenge is to completely remove phenomenological aspects
in EDF and directly connect terms entering in the functional to the parameters of the bare interaction derived 
from chiral EFT. Doing so gives the bottom-up picture displayed in figure \ref{fig:pairinggap} (left) and promotes 
the EDF framework to the rank of an ab initio theory. A discussion on the strategies to design ab-initio EDF taking advantage 
of modern nuclear forces specificities can be found in ref. \cite{Dru10}. Conjointly to this long term project, similarly 
to the Shell-Model case, attempts to combine conventional EDF with modern forces are being now developed. An illustration of
the pairing gap obtained when the $V_{\rm lowk}$ interaction (and Coulomb for protons) 
is used in the pairing channel of the functional while keeping the mean-field channel unchanged (Skyrme force) \cite{Dug09b}.     
Although the possibility to combine effective interaction and bare interaction into a single calculation is still to 
be clarified, the calculated gap with bare interaction is in very good agreement with experimentally known nuclei. 
In particular, the prediction is much better than the empirical predictions based on phenomenological contact pairing 
interactions. Future experiments with radioactive nuclei and the possibility to add more points on the neutron rich side 
will be crucial to further validate such calculations.     

\section{Summary}

\begin{figure}[htb]
\begin{center}
\includegraphics[width = 17.cm]{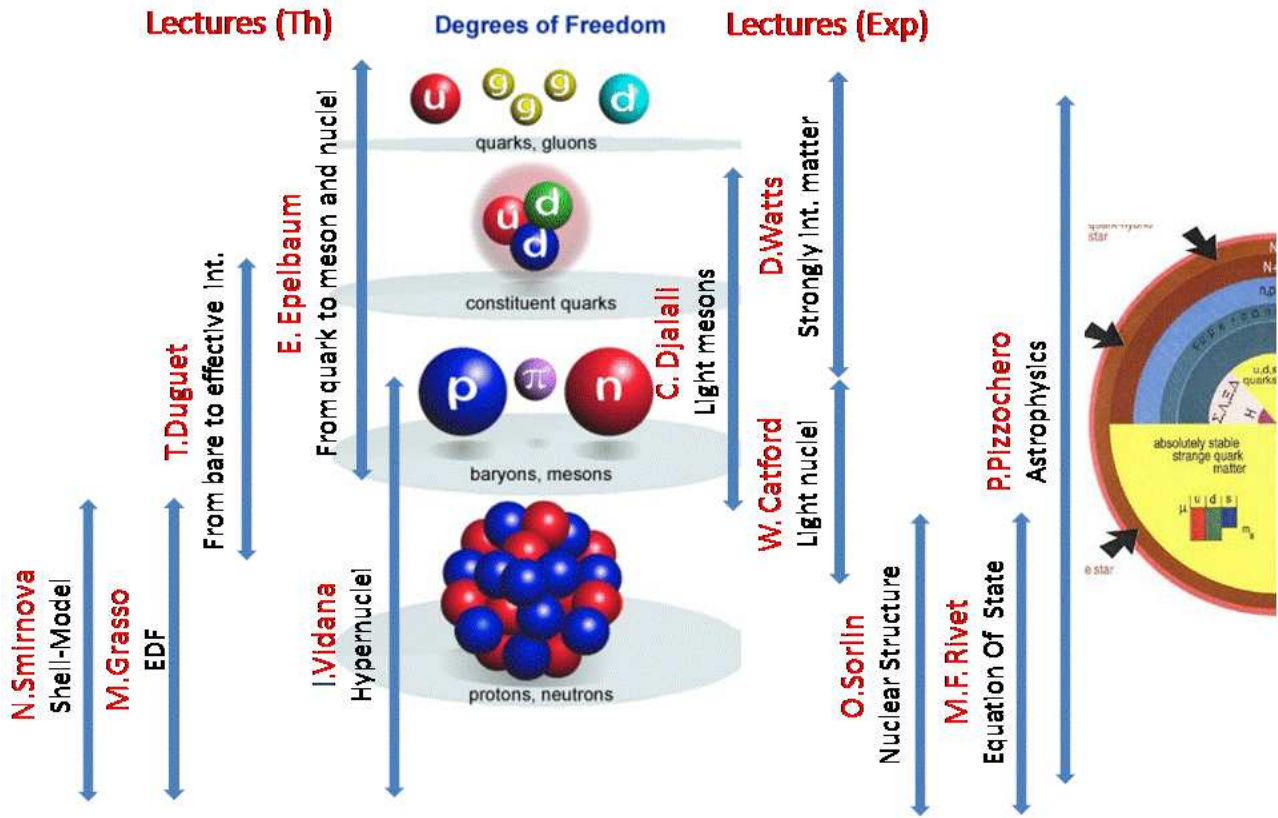}
\end{center}
\caption{(Color online) Schematic illustration of the different lectures given in the 2009 International Joliot-Curie School. 
Lectures displayed in the left and right side respectively correspond to more theoretical or  more experimental.
}
\label{fig:lectures}
\end{figure}

In this introduction to the 2009 International Joliot-Curie School (EJC2009), recent advances related 
to nuclear interaction have been summarized starting from quarks and ending with finite or infinite nuclear 
systems. At different energy scale, selected new concepts and ideas have been discussed in a rather
simple, sometimes oversimplified, way. Detailed discussion on the many facets of nuclear interaction 
can be found in dedicated lecture of this school. A road map of the different EJC2009 lectures as well as 
physical features each lecture is covering is given in figure \ref{fig:lectures}. I hope that this introduction 
will be helpful to provide a coherent picture for the school.   

Recent achievement in nuclear forces theory open new perspectives for the next decade 
of low energy nuclear physics, bringing together people from very different communities. Although many developments 
remain to be done,   
the possibility to directly use QCD to describe nuclear system is a major challenge that is within reach.    

\begin{center} 
{\bf Acknowledgments}
\end{center}
I thank the organizing committee and different lecturers of the EJC2009 school for helpful discussions 
and for providing some of their transparencies to prepare this introduction. The author thank 
G. Hupin and G. Burgunder for proofreading the manuscript.

\bibliography{biblio_abc,biblio_def,biblio_ghi,biblio_jkl,biblio_mno,biblio_pqr,biblio_stu,biblio_vwxyz}
\end{document}